\documentclass[showkeys,showpacs,prc,nofootinbib]{revtex4}
\usepackage{exscale}
\usepackage[dvips]{graphicx}

\newcommand{\bea}{\begin{eqnarray}}
\newcommand{\eea}{\end{eqnarray}}
\newcommand{\p}{$\phi$ }
\newcommand{\bc}{\begin{center}}
\newcommand{\ec}{\end{center}}
\newcommand{\bfi}{\begin{figure}}
\newcommand{\efi}{\end{figure}}
\newcommand{\igr}{\includegraphics}

\begin{document}
\title{Photoproduction of $\phi$ mesons from nuclei}
\thanks{Supported by DFG and BMBF}
\author{P. M\"uhlich}
\email{pascal.muehlich@theo.physik.uni-giessen.de}
\author{T. Falter}
\author{C. Greiner}
\author{J. Lehr}
\author{M. Post}
\author{U. Mosel}
\affiliation{Institut f\"ur Theoretische Physik, Universit\"at Giessen, D--35392 Giessen, Germany}

\keywords{photonuclear reactions, $\phi$ decay in nuclei, medium properties of mesons}
\pacs{25.20.Lj, 13.25.-k, 14.40.Cs}

\begin{abstract}
We investigate the consequences of possible medium modifications of the $\phi$ meson at finite nuclear matter density on the $K^+K^-$ mass distribution in photonuclear reactions. The inclusive cross sections for $K^+K^-$ pair production are calculated within a semi--classical BUU transport model, which combines the initial state interaction of the incoming photon with the final state interactions of the produced particles. The effects of final state interactions on the invariant mass distribution of the observed $K^+K^-$ pairs are discussed in detail. In addition we calculate the Coulomb correction and possible effects of hadronic kaon potentials on the $K^+K^-$ mass spectrum. Due to the large cross sections for reactions of the final state particles with the surrounding nuclear medium and the influence of the Coulomb potential we find no measurable sensitivity of the observables to the \p properties at finite baryon density.
\end{abstract}

\maketitle

\section{Introduction}\label{introduction}

Medium modifications of hadrons result on the one hand from standard hadronic many--body effects, which first of all appear to be completely independent of the partonic sub--structure of these particles. In particular the determination of the $\rho$ meson spectral function has attracted much theoretical interest in recent years \cite{asakawa,rapp,post}. On the other hand, it is expected, that the properties of hadrons are influenced by the spontaneous breaking of chiral symmetry \cite{brown}. The authors of \cite{hatsuda,kampfer,Leupold:zh} calculated within a QCD sum rule approach, that the masses of the light vector mesons drop in nuclear matter due to a partial restoration of chiral symmetry and the lowering of the chiral condensate. Up to now it is completely unknown, in which manner the mentioned many-body effects are connected with chiral symmetry restoration \cite{koch}.

In this respect also the in--medium properties of the $\phi$ meson have been a subject to several theoretical studies \cite{weise,klingl,ramos}.  The predictions of these works have in common a quite small mass shift at normal nuclear matter density but a sizeable enhancement of the $\phi$ width up to one order of magnitude above its free width of 4.4 MeV. The investigation of the $\phi$ in--medium properties is of great interest due to various reasons: On the one hand the $\phi$ is neither in the dilepton nor in the $K\bar K$ invariant mass spectrum concealed by other resonances. On the other hand the $\phi$ properties are of fundamental interest for basic features of QCD, since one expects to gain not only information about chiral symmetry restoration but also about the strangeness admixture in the nucleon \cite{oh} and the appearance of the quark--gluon plasma in heavy--ion collisions \cite{shor,chung}.

Furthermore, there has been the experimental effort to look for the existence of a small, but significant decrease in the peak mass of the invariant mass spectrum of the \p meson, occuring for central heavy ion collisions. However, these measurements found that the actual \p mass distribution was consistent with that of the free \p meson \cite{akiba}.

The purpose of the present paper is to analyse in detail the feasibility of an experiment running soon at Spring8/Osaka \cite{osaka}. The considered reaction is inclusive photoproduction of $\phi$ mesons from nuclei, looking at the $K^+K^-$ invariant mass spectrum, which contains both information about the mass and the width of the produced $\phi$ mesons. The experiment will be performed with a photon beam of 2.4 GeV maximum energy, which is produced by backward Compton scattering of laser photons from an 8 GeV electron beam. As pointed out in \cite{osaka}, the estimated mass resolution of the detector is 600 keV, which is much smaller than the free $\phi$ width. Thus, the observables could be highly sensitive to small deviations of the mass distribution compared to the situation in free space.

Already the authors of \cite{oset} performed a many--body calculation with a similar purpose as in the present paper. In this study the photon self--energy diagram for a $\phi$ and a particle--hole in the intermediate state was calculated. They used the results for the $\phi$ width obtained in \cite{ramos}, which contains 'elastic' contributions  ($\phi\rightarrow K\bar K$) due to the renormalization of the kaons and 'inelastic' contributions due to $\phi N$ absorption reactions ($\phi N\rightarrow K \Lambda,\Sigma$, etc.). The final state interactions (FSI) were taken into account by a Glauber--type geometrical damping factor for outgoing anti--kaons. As a final result the authors of \cite{oset} found an observable enhancement of the width of the invariant $K^+K^-$ mass distribution by a factor of about two.

At first sight, the observation of medium properties of the $\phi$ through the $K\bar K$ invariant mass spectrum seems rather unlikely due to two reasons: First, most of the $\phi$ mesons will decay outside the nucleus due to the small width of the $\phi$ resonance. As will be discussed below, this problem can be avoided by restricting the three momenta of the observed kaon--anti--kaon pairs to about 150 MeV as already proposed in \cite{oset}. Second, the invariant mass spectrum might be strongly distorted by $KN$ and $\bar KN$ reactions, for which the cross sections are quite large. The influence of these effects on the shape of the mass spectrum will be discussed in detail in the section concerning the results of our calculations.

For the description of the photonuclear reaction we make use of a semi--classical BUU transport model. Our model allows the calculation of inclusive particle production in heavy ion collisions from 200 AMeV to 200 AGeV \cite{teis} and in pion \cite{brat}, photon \cite{effe96,effe00} and electron \cite{lehr} induced reactions with just the same physical input. Already in \cite{effe99} observable consequences of medium modifications of the light vectormesons $\rho$ and $\omega$ through their dileptonic decay in photon--nucleus reactions were investigated. A presentation of the model in full detail was given in \cite{effe96}.

In the BUU model the photon--nucleus reaction is described as a two--step process. In the first step the incoming photon reacts with a single nucleon (impulse approximation). Since for the considered photon energy the effect of nuclear shadowing has an observable impact on the cross sections \cite{falter_schatt}, the shadowing effect is taken into account as explained in \cite{falter_form}. In the second step the produced particles are propagated explicitly through the nucleus allowing for all kinds of final state interaction. This part of our model is governed by the semi--classical transport equations.

We take into account elastic \p photoproduction as well as inelastic \p production as elementary $\gamma N$ processes. An important ingredient to our model is the elastic \p photoproduction cross section, for which various models are available  \cite{landshoff,laget,titov,kochelev}. Moreover, inelastic \p photoproduction, i. e. processes of the type $\gamma N\rightarrow \phi\pi N$ and $\gamma N\rightarrow \phi K \bar K N$, turns out to play the dominant role in the regime of \p momenta below a few hundred MeV and therefore has to be included in order to allow calculations yielding absolute numbers to be compared to experiment. Since for these processes neither experimental data nor microscopic models are available, we simulate these events using the Lund model FRITIOF \cite{fritiof}.

The inelastic \p photoproduction processes have not been included in \cite{oset}. Furthermore, compared to the purely absorptive FSI of \cite{oset} our model allows for a more realistic description of the FSI, taking into account also regeneration, side--feeding and inelastic scattering of the considered mesons. Inelastic $\phi N$ scattering, i. e. $\phi N\rightarrow \phi X$, yields a slowing down of \p mesons, which is particularly important when the above mentioned momentum cuts are applied. As in the case of inelastic \p photoproduction, we use the FRITIOF routine to model these reactions. Also the distortion of the mass spectrum by elastic FSI can be investigated directly in our approach due to the explicit inclusion of elastic scattering processes.

Our paper is organized as follows: In section \ref{xsection} we discuss briefly the elementary cross sections for \p photoproduction from the nucleon. In section \ref{buu} we describe the BUU equation and the cross sections concerning the FSI. Some details of our calculations and the final results, studying thoroughly the visibility of medium modifications of the $\phi$, are presented in section \ref{results}. We close with a summary in section \ref{summary}.


\section{Elementary $\phi$ photoproduction cross sections}\label{xsection}

\subsection{Inelastic $\phi$ photoproduction}

For the simulation of high energy photoproduction processes we use the Lund model FRITIOF \cite{fritiof}. In the FRITIOF model particle production occurs as a two step process. First, two excited states with the quantum numbers of the incoming hadrons are produced (string excitation). In the second step these strings fragment into observable hadrons (fragmentation).

Since the FRITIOF model does not accept photons in the entrance channel, we use the vector meson dominance model to replace the photon by a massless $\rho$, $\omega$ or $\phi$ meson. The relative probability $p_V$ to initialize a particular vector meson $V$, i. e. the hadronic content of the incoming photon, is given by the strength of the coupling of the vector meson to the photon times its nucleonic cross section $\sigma_{VN}$:
\bea
p_V=\left(\frac{e}{g_V}\right)^2\frac{\sigma_{VN}}{\sigma_{\gamma N}},
\eea
where $\sigma_{\gamma N}$ is the total photon nucleon cross section. For a detailed review of this method we refer to \cite{falter_form}.

In \cite{geiss} it has been checked, that the FRITIOF model gives a good description of inclusive particle production for elementary hadronic reactions. However, one cannot expect to get an equally good description of elastic processes. Therefore elastic vector meson production as well as exclusive strangeness production \cite{effe00} are treated independent of the FRITIOF model, if explicit cross sections are available.

The only nucleonic resonance that FRITIOF produces as a product of the fragmentation process is the $\Delta(1232)$. As a consequence of the missing higher nucleon resonances there might be too much strength in outgoing channels including the $\Delta$ resonance. This can be seen in the angular differential cross section for the reaction $\gamma N \rightarrow \phi X$, which due to the particular process $\gamma N\rightarrow \phi \Delta$ exhibits an unphysical peak in forward direction.

Therefore we treat not only the elastic $\phi$ production process but also the process $\gamma N\rightarrow\phi\Delta$ independent of FRITIOF. We parameterize the total cross section with the following ansatz:
\bea
\label{sigtot}
\sigma=\frac{1}{p_i s}\int d\mu_{\Delta}S_{\Delta}(\mu_{\Delta})|\mathcal{M}|^2p_f(\mu_{\Delta}),
\eea
where $\mathcal{S}_{\Delta}$ denotes the spectral function of the $\Delta$. For the angular distribution we use an exponential form factor times a propagator, assuming a dominant contribution from $t$--channel exchange:
\bea
\frac{d\sigma}{dt}\propto \exp (Bt)\left(\frac{1}{t-m_{\pi}^2}\right)^2
\eea
For the parameter $B$ we take a value of $B=6.5$ GeV$^{-2}$ from a fit to inelastic $\phi$ photoproduction data in the mass region of the $\Delta$ resonance in ref. \cite{desy}. We adjust the constant matrix element for the total cross section in order to describe the experimental data points for inelastic $\phi$ photoproduction in the mass range $1.2~\mathrm{GeV}\le M_X <2.1~\mathrm{GeV}$, measured at photon energies between 4.85 GeV and 5.85 GeV \cite{desy}. We find a value of $\mathcal{M}=0.0866$. In figure \ref{tsigma_inc} we show the resulting inelastic $\phi$ photoproduction cross section in comparison with experimental data at different energies. The agreement is quite good at low energies, whereas the data are less well reproduced at higher energies.


\subsection{Elastic $\phi$ photoproduction}\label{exsection}

For the elastic $\phi$ photoproduction process we use model A from ref. \cite{titov}. In this model the $\phi$ photoproduction amplitude is given as a coherent sum over the pomeron--,  $\pi$--, $\eta$-- and $s$-- and $u$--channel exchange diagrams.

In principle the $\phi NN$ coupling in the Born diagrams should be dressed by a cutoff form factor for off--shell nucleons. Using a $s$--dependent form factor for the $s$--channel and a $u$--dependent form factor for the $u$--channel would result in a violation of gauge invariance with respect to the photon and $\phi$ meson fields alike. Hence in \cite{titov} a constant form factor $F=F_s=F_u=1$ was used. This leads to an overestimation of the angular differential cross section for high momentum transfers, where the $s$-- and $u$--channel contributions become dominant.

Following the model of ref. \cite{haberzettl} we therefore introduce an overall form factor, which keeps the gauge invariance of the amplitude as well: 
\bea
\left(\mathcal{M}_s+\mathcal{M}_u\right)'=F(s,u)\left(\mathcal{M}_s+\mathcal{M}_u\right).
\eea
This form factor is composed as follows:
\bea
F(s,u)=a_s F(s)+a_u F(u)\quad \mathrm{with} \quad F(x)=\Lambda^4/(\Lambda^4+(x-m_N^2)^2),
\eea
where the prefactors $a_s$ and $a_u$ have to fulfill the constraint $a_s+a_u\equiv 1$. We fit the parameters of the model in order to describe the data at a photon energy of 2 GeV \cite{be74} as well as the data at 3.6 GeV \cite{ac00}. Excellent agreement can be achieved with a cutoff of $\Lambda=1.87$ GeV and factors $a_s=0.9$ and $a_u=0.1$. The main effect of this form factor then is the supression of the $s$-- and $u$--channel contributions at large values of the invariant energy $s$. The resulting cross section is shown in figure \ref{tsigma_exc} in comparison with the experimental data.

As already mentioned in the introduction one has to apply severe momentum cuts to the $K^+K^-$ three momentum in the $\gamma A$ reaction in order to guarantee that a considerable fraction of events stems from $\phi$ decays inside the nucleus. In figure \ref{psigma_exinc} we compare the momentum spectrum of exclusive (=elastic) and inclusive (=elastic+inelastic) produced $\phi$ mesons. One can see that in the region of low $\phi$ momenta the cross section is dominated by the inelastic $\phi$ photoproduction processes by about one order of magnitude. Therefore it is essential to take the inelastic processes into account in order to investigate the sensitivity of observables on the in--medium properties of the $\phi$.


\section{The transport model}\label{buu}

Since our model was presented extensively in \cite{teis,brat,effe96,effe00,lehr,effe99} and references therein, we restrict ourselves to a brief description of some essential features of the model, which are of immediate interest to the present work.

\subsection{The BUU equation}

The BUU equation describes the time evolution of the one--particle phase--space density $f(\vec r,\vec p, t)$ under influence of an effective mean field potential $U(\vec r,\vec p,f)$ and a collision term taking account of the Pauli principle. For the description of a system of particles with continous mass spectra the classical equation is extended by defining the spectral phase--space density as the product of the ordinary phase--space density $f$ and the particle spectral function $S$ \cite{effe99,um}:
\bea
F(\vec r,\vec p,\mu ,t)=f(\vec r,\vec p,t)S(\vec r,\vec p,\mu ,t),
\eea
where $\vec r$ and $\vec p$ are the spatial and momentum coordinates of the particle and $\mu$ denotes its invariant mass. The set of transport equations for a system of $N$ particle species then is given by
\bea
\label{eom}
(\partial_t+\partial_{\vec p}\mathcal{H}_i\partial_{\vec r}-\partial_{\vec r}\mathcal{H}_i\partial_{\vec p})F_i=I_i[F_1,...,F_N]\qquad i=1,...,N
\eea
where $I_i[F_1,...,F_N]$ denotes the collision term, which now also contains the spectral information of the particle. $\mathcal{H}_i=\mathcal{H}_i(\vec r,\vec p,\mu ,F_1,...,F_N)$ is the single--particle mean field Hamilton function, which in our model is given by the relativistic expression for the single--particle energy
\bea
\mathcal{H}_i=\sqrt{(\mu_i+U_i)^2+\vec p_i^{~2}},
\eea
with the effective scalar mean field potential $U_i=U_i(\vec r,\vec p,\mu,F_1,...,F_N).$ 


\subsection{The collision term}

The collision term can be decomposed in a gain and a loss term. In an obvious notation the collision term then reads:
\bea
I_i(\vec r,\vec p,\mu ,F_1,...,F_N)=G_iS_i(1\pm f_i)-L_iF_i,
\eea
where the upper sign holds for bosons and the lower sign for fermions, in the latter case accounting for the Pauli principle. 

The gain term accounts for the creation of particles of the type $i$ in the phase--space element with coordinates $(\vec r,\vec p,\mu )$ due to all kinds of collisions with particles of type $i$ in the final state. Rewriting the general expression for $G_i$ in terms of cross sections for these processes, neglecting $n$--body collisions, $n\ge 3$, one obtains
\bea
G_i=(2\pi)^3\sum\limits_{1,2}\sum\limits_{X}\int\frac{d^3p_1d\mu_1}{(2\pi)^3}\frac{d^3p_2d\mu_2}{(2\pi)^3}v_{12}F_1F_2\left(\frac{d^3\sigma}{dp_i^3}\right)_{12\rightarrow iX} (1\pm f_3)...(1\pm f_m)
\eea
where $v_{12}$ is the relative velocity of the incoming particles 1 and 2 and $X$ denotes a final state consisting of the particles $3,...,m$.

The loss term accounts for the annihilation of particles of type $i$ and can be written in the following way:
\bea
L_i=\Gamma_{i\rightarrow X}^*+\sum\limits_2\rho_2\langle\sum\limits_X v_{i2}\sigma_{i2\rightarrow X}(1\pm f_3)...(1\pm f_m)\rangle,
\label{loss}
\eea
where $\Gamma_{i\rightarrow X}^*$ denotes the total decay width of the particle species $i$ including medium corrections, i. e. Pauli blocking / Bose enhancement of the final state. $\rho_2$ is the spatial density of the particle species 2 and $v_{i2}$ is again the relative velocity of the incoming particles. The square brackets have to be understood as an average over the momentum distribution of particle species 2. Note that the second term of expression (\ref{loss}) has the general form of a collision width, arising from two--body collisions of particle species $i$ and 2. 

\subsubsection{$\phi N$--collisions}\label{phin}

For invariant energies below $\sqrt{s}=2.2$ GeV we include the following processes for $\phi$--nucleon collisions:
\begin{eqnarray*}
\qquad\qquad\qquad\qquad\qquad\qquad\qquad\phi N &\rightarrow &\phi N\\
\phi N &\rightarrow &K\Lambda\\
\phi N &\rightarrow &\pi N\\
\phi N &\rightarrow &\phi\Delta, \eta N, \eta\Delta, \eta'N, \eta'\Delta, K\bar KN, \pi\pi N
\end{eqnarray*}
For the elastic scattering process we exploit the vector meson dominance model (VMD). We take the model for elastic $\phi$ photoproduction, which we discussed in the previous section \ref{exsection}, and replace in the appropriate diagrams the incoming photon by an incoming $\phi$, using a coupling constant $g_{\phi}^2/4\pi=13.2$ from \cite{emc}. The $\pi$--exchange diagram does not contribute to elastic $\phi N$ scattering, since there is no $\phi\phi\pi$--coupling due to isospin conservation. This method also delivers the angular distribution for elastic $\phi N$--scattering. We use this cross section also for invariant energies above 2.2 GeV.

For the process $\phi N\rightarrow K\Lambda$ we adopt the cross section of ref. \cite{ko}, which has been calculated based on a kaon exchange model. The cross section for the process $\phi N\rightarrow \pi N$ we obtain via the detailed balance relation
\bea
\sigma_{\phi N\rightarrow\pi N}=\frac{1}{3}\left(\frac{p_f}{p_i}\right)^2\sigma_{\pi N\rightarrow\phi N},
\eea
where $p_f$ is the momentum of the $\pi N$ system and $p_i$ the momentum of the $\phi N$ system. The cross section $\sigma_{\pi N\rightarrow \phi N}$ we take from \cite{sibirtsev}, where a resonance parameterization was fitted to experimental data.

For the cross sections of the remaining processes we make a simple phase--space model. The cross sections of processes involving a $\Delta$ resonance in the final state are parametrized by an ansatz similar to equation (\ref{sigtot}). For the processes $\phi N\rightarrow \eta N$ and $\phi N\rightarrow \eta' N$ we assume the final state particles to be sharp states. Hence we have
\bea
\sigma=\frac{p_f}{p_i s}|\mathcal{M}|^2.
\eea
Finally we take the cross sections for processes with three particles in the final state as constant for simplicity. The matrix elements of all these cross sections are fitted in order to allow for a smooth transition to the high energy cross sections provided by the FRITIOF model, which we exploit for energies above $\sqrt{s}=2.2$ GeV, see section \ref{fritzi}. The cross sections discussed above are displayed in figure \ref{sigma_low}. We also note that all these cross sections are not accessible by experiment and therefore a considerable uncertainty arises. We will come back to this point when discussing the results of our calculations.
 
\subsubsection{$KN$-- and $\bar KN$--collisions}

For collisions of kaons and nucleons at energies below $\sqrt{s}=2.2$ GeV we include the following processes:
\begin{eqnarray*}
KN & \rightarrow & KN\\
KN & \rightarrow & K\pi N.
\end{eqnarray*}
For anti--kaon--nucleon collisions we have the following processes:
\begin{eqnarray*}
\bar K N & \rightarrow & \bar K N\\
\bar K N & \rightarrow & \pi \Lambda\\
\bar K N & \rightarrow & \pi \Sigma\\
\bar K N & \rightarrow & \pi Y^*.
\end{eqnarray*}
The cross sections for these processes are discussed in detail in reference \cite{effe00}, appendix A. In contrast to the $\phi N$ cross sections the cross sections for $KN$ and $\bar KN$ collisions are well confirmed by experimental data \cite{effe00}.

\subsubsection{FRITIOF}\label{fritzi}

For meson--baryon collisions at invariant energies above $\sqrt{s}=2.2$ GeV we use the Lund  model FRITIOF \cite{fritiof}, which we sketched already in the section dealing with the inelastic photoproduction channels. In the FRITIOF model hadrons are described in the simple valence--quark picture, i. e. baryons consist of three quarks whereas mesons are build of a quark and an antiquark. Since FRITIOF does not take into account quark exchange between the two initial hadrons, some processes are not possible, for instance the process $\pi N\rightarrow\Lambda K$ (cf. ref. \cite{geiss}). Such flavor exchange processes are important primarily at low energies.

As already mentioned earlier, one cannot expect to get a good description of elastic scattering processes by the FRITIOF model. Therefore, we treat elastic vector meson--nucleon scattering for all energies independent of the FRITIOF model.

The FRITIOF model only delivers the relative abundances of the various hadronic final states. In order to obtain cross sections the total cross section for a given initial state has to be put in by hand. For high energy $\phi$--nucleon collisions we use a total cross section of $\sigma_{\phi}=12$ mb, which is motivated by $\phi$ meson photoproduction at high energies using the vector meson dominance model \cite{bauer}. In figure \ref{sigma_high} we show the resulting high energy $\phi$--nucleon cross sections. 


\subsection{The $\phi$ spectral function}

The $\phi$ spectral function is in our model parameterized by a relativistic Breit--Wigner distribution
\bea
\label{spectral}
S_{\phi}(s)=\frac{2}{\pi}\frac{s\Gamma_{\mathrm{tot}}(s)}{(s-m_0^2)^2+s\Gamma_{\mathrm{tot}}^2(s)},
\eea
where $\Gamma_{\mathrm{tot}}(s)$ denotes the total energy--dependent width and $m_0$ the pole mass of the \p. For the energy dependence of the hadronic decay widths of all baryonic and mesonic resonances we adopt the parameterization of \cite{manley}, which has also been described in \cite{effe99}. For the decay of a $\phi$ meson into a $K\bar K$--pair at energies around the $\phi$ meson pole mass the general expression for the width can be reduced to  
\bea
\label{decwidth}
\Gamma_{\phi\rightarrow K\bar K}(\mu)\approx \Gamma_{\phi\rightarrow K\bar K}(m_0) \frac{m_0}{\mu}\left(\frac{p(\mu)}{p(m_0)}\right)^3,
\eea
where $\mu$ is the invariant mass of the $\phi$ and $p(\mu)$ is the kaon momentum in the $\phi$ rest frame. This expression has exactly the same energy dependence as the $\phi$ width derived by a one--loop calculation  in \cite{klingl} and \cite{ramos}.

For the in--medium width of the $\phi$ we make the following ansatz:
\bea
\label{ansatz}
\Gamma_{\mathrm{med}}(\vec r,|\vec p|,\mu)=\Gamma_{\mathrm{vac}}(\mu)+\Gamma_{\mathrm{coll}}(\vec r,|\vec p|,\mu)
\eea
with the vacuum width $\Gamma_{\mathrm{vac}}$, which is a sum of the $K\bar K$ width as discussed above and a rather small contribution from the $\rho\pi$--channel, and a collision width $\Gamma_{\mathrm{coll}}$, which in addition depends on the spatial and momentum coordinates of the $\phi$. According to equation (\ref{loss}) the collision width due to $\phi N$ collisions is given by
\bea
\Gamma_{\mathrm{coll}}(\vec r,|\vec p|,\mu)=\gamma\rho\langle v_{\mathrm{rel}}\sigma_{\phi N}^{\mathrm{tot}}\rangle,
\label{collwidth}
\eea
where $\gamma$ denotes the Lorentzfactor for the transformation from the nuclear rest frame to the $\phi$ rest frame. The brackets indicate an average over the Fermi motion of the nucleons, $v_{\mathrm{rel}}$ is the relative velocity of nucleon and meson and $\sigma_{\phi N}^{\mathrm{tot}}$ is the total $\phi N$ cross section. At nuclear density $\rho_0=0.17$ fm$^{-3}$ and vanishing $\phi$ momentum the $\phi$ collision width then amounts to 24 MeV and therefore is comparable to the $\phi$ width obtained by a more involved many--body approach in \cite{ramos}. We also note that by means of the ansatz (\ref{ansatz}) the decay width of the \p into $K\bar K$--pairs is not modified, since the complete collision width corresponds to other decay channels, for instance $\phi N\rightarrow K\Lambda$\footnote{This holds as long as the masses of $K$ and $\bar K$ are not modified in the medium. Including also a $K/\bar K$ renormalization the modification of the decay width is obviously given by the momentum dependence of expression (\ref{decwidth}).}.


\subsection{Off--shell transport}

The transport equations (\ref{eom}) alone do not yield the right asymptotic phase--space densities, since a particle, which is collisional broadened in the medium, will not automatically lose its width as it propagates into the vacuum\footnote{Actually, particles with large width are driven back to their free width by the collision term alone, see \cite{effe99}.}. Therefore an ad--hoc method was introduced in \cite{effe99,um} in order to allow for a dynamical treatment of particle mass spectra.

Within this approach the off--shellness of a given particle is absorbed in a scalar potential, which is defined as
\bea
s_i(\rho_i(t))=(\mu_i^{\mathrm{med}}-\mu_i^{\mathrm{vac}})\frac{\rho_i(t)}{\rho_i(t_{\mathrm{cr}})},
\eea
where $\mu_i^{\mathrm{med}}$ denotes the in--medium mass of the particle, choosen according to the in--medium spectral function, and $\mu_i^{\mathrm{vac}}$ denotes the particle mass which is choosen according to the vacuum spectral function. $\rho_i(t)$ is the baryon density during the propagation at time $t$, whereas $\rho_i(t_{\mathrm{cr}})$ is the density at the creation point. The effective in--medium mass of the particular particle is given by:
\bea
\label{mstar}
\mu_i^*(\rho_i(t))=\mu_i^{\mathrm{vac}}+s_i(\rho_i(t)).
\eea
This description obviously yields an evolution of the particle mass to its correct asymptotic value. At the creation point inside the nucleus the particle mass is equal to the in--medium mass $\mu_i^*(\rho(t_{\mathrm{cr}}))=\mu_i^{\mathrm{med}}$, whereas as the particle travels outside the nucleus, it propagates back to a mass according to its vacuum spectral function $\mu_i^*(\rho=0)=\mu_i^{\mathrm{vac}}$. To guarantee energy conservation, the potential $s_i$ enters as an usual scalar potential into the equations of motion.

This recipe has been justified a posteriori by the works of \cite{leupold} and \cite{juchem}. Their more refined transport equations go over into our off--shell transport description by assuming that the particle width scales linearly with the density, neglecting its momentum dependence, and that the off--shellness (i. e. $E_i-E_0$, where $E_i$ is the particle energy during the propagation and $E_0$ is the appropriate on--shell energy) is small. Both constraints are fulfilled quite well by the \p meson, which gains a collision width in the medium. Since the \p width is small compared to its mass, the off--shellness indeed should be small. A collision width, which arises from two body collisions of the \p meson with nucleons in the nuclear environment, has in first order a linear dependence on the nuclear matter density (see equation (\ref{collwidth})).


\section{Results}\label{results}

We perform our calculations for a $^{40}$Ca nucleus at a photon energy of 2.4 GeV, which is the maximum beam energy of the Osaka experiment \cite{osaka}. We initialize the nucleons in $\vec r$--space according to the Woods--Saxon density distribution with the parameters given in \cite{lehr}, whereas the distribution of protons and neutrons is assumed to be the same. For the initialization in momentum space we apply the local density approximation, where the ground state nucleon momenta are distributed homogeneously in the Fermi sphere up to a local Fermi momentum $p_F(r)$, given by its nuclear matter value at density $\rho(r)$: 
\bea
p_F(\vec r)=\left(\frac{3}{2}\pi^2\rho(\vec r)\right)^{1/3}.
\eea

As mentioned earlier the $K^+K^-$ three momentum has to be restricted to very small values in order to ensure that a sufficient fraction of events stem from \p decays at finite density. From the life--time and the velocity of a given \p meson one can easily get an estimate on the distance $d$ which the \p has travelled during its life--time (i. e. the time after which the \p has decayed with a probability of 63\%), which has to be compared to the radii of the considered nuclei. Corresponding to the radius of $^{40}Ca$, $R\approx 4$ fm, we use three different cutoff values $\Lambda$ for the \p three--momentum: $\Lambda=300$ MeV ($d=12.8$ fm), $\Lambda=150$ MeV ($d=6.4$ fm) and $\Lambda=100$ MeV ($d=4.3$ fm).

The observable of interest, which eventually exhibits modifications due to an in--medium change of the \p mass and width, is the mass differential cross section for inclusive photoproduction of $K^+K^-$--pairs from nuclei. Within the transport approach we write the mass differential photon--nucleus cross section in the following way:
\bea
\label{sigma}
\left(\frac{d\sigma}{dM}\right)_{\gamma A\rightarrow K^+K^- X} & \sim & \int d^3r\int\limits^{p_F(\vec r)}d^3p\int d\mu~\Theta(\Lambda-|\vec p_{K^+}+\vec p_{K^-}|)\delta(M-m_{K^+K^-})\nonumber\\ & & \qquad\qquad
\times\left\{\sigma_{\gamma N\rightarrow\phi X}S_{\phi}(\mu)\frac{\Gamma_{\phi\rightarrow K^+K^-}}{\Gamma_{\phi}^{\mathrm{tot}}} + \left(\frac{d\sigma}{d\mu}\right)^{\mathrm{non-res}}_{\gamma N\rightarrow K^+K^-X}\right\}\mathcal{N}_{K^+K^-},
\eea
where $\mu$ is the invariant mass of the primary produced \p meson or $K^+K^-$--pair and $m_{K^+K^-}$ is the asymptotic invariant mass after the propagation. $S_{\phi}(\mu)$ denotes the \p spectral function and $\Lambda$ is the momentum cutoff parameter. The factor $\mathcal{N}_{K^+K^-}$ specifies the asymptotic number of $K^+K^-$--pairs with invariant mass $m_{K^+K^-}$, which is the result of the solution of the coupled system of transport equations. 

At photon energies above around one GeV also the effect of nuclear shadowing, which, for simplicity, we did not account for in equation (\ref{sigma}), is not neglegible \cite{falter_schatt}. The implementation of this coherence effect into a semi--classical transport model is described in detail in \cite{falter_form}. In our calculations the effect of nuclear shadowing yields a reduction of the cross sections by about 15\%.


\subsection{Inclusive \p photoproduction from nuclei}\label{phiproduction}

\subsubsection{Momentum spectrum}

First we consider the momentum differential cross section for \p photoproduction from $^{40}$Ca, depicted in figure \ref{psigma_fsi}. In this plot we show five different curves, corresponding to different scenarios: two curves belong to calculations in which only elastic $\phi$ production is taken into account as elementary $\gamma N$--process and two curves correspond to calculations where also inelastic \p production is taken into account. For both cases we show calculations with and without final state interactions (FSI). Furthermore we also show the momentum spectrum for the total inclusive $K^+K^-$--photoproduction reaction, including also non--resonant background production of $K^+K^-$--pairs (i. e. the reaction $\gamma N\rightarrow K^+K^- N$ which processes not via an intermediate \p meson).

Comparing the curves of the exclusive calculations, which in the case without FSI drop very rapidly to low momenta, we find an appreciable enhancement of the spectrum in the region of low \p momenta when the FSI are included. This modification arises in the first place from inelastic $\phi$--nucleon reactions, i. e. the processes $\phi N\rightarrow \phi X$. The final state $X$ mainly consists of a nucleon or a nucleon resonance and one or several pions. The inclusive calculations show a similar behavior, albeit it is not as obviously visible as in the mere exclusive case. This is due to the absorption of primarily slow \p mesons produced by inelastic photoproduction processes, which acts against the production of low--momentum \p mesons during the FSI. Even though the inclusive and exclusive curves differ at intermediate momenta by up to one order of magnitude in calculations without FSI, the curves with FSI differ in the region of interest at momenta below 300 MeV merely by a factor of about two.

Due to this slowing down we find a strong sensitivity of the calculated observables on the $\phi$--nucleon cross sections. As already mentioned in section \ref{phin} these cross sections are not accessible by experiment and therefore a considerable uncertainty arises. Hence we perform two additional calculations where we multiply/divide all cross sections for $\phi$--nucleon processes by a factor of two in order to get an estimate on the systematic uncertainties of our calculations. The results are shown in figure \ref{psigma_enh}. In the region of low \p momenta we find discrepancies of the cross sections in both directions of up to 30\%. Note that an increase of the $\phi N$ cross sections yields a further enhancement of the $\phi$ momentum spectrum at very low momenta, stressing again the more significant influence of the inelastic $\phi N$ scattering processes, which enhance the slow \p meson yield, compared to the absorption of \p mesons at low momenta. Since we artificially put in an extreme modification of the \p in--medium cross sections, which -- at least for high \p energies -- can be excluded by the VMD relation to high energy \p photoproduction, we can be quite confident that the systematic error of our calculations due to the $\phi$--nucleon cross sections is not larger than the estimated value.

The elastic and inelastic $\phi$--nucleon reactions increase the yield of low--mo\-mentum \p mesons by a large factor. Thus, most of the slow \p mesons, which are responsible for in--medium effects in the observables, are produced by rescattering. For this reason these processes are certainly not negligible. This again underlines the advantage of our model compared to more simple Glauber--type models as \cite{oset}, which treat the FSI as purely absorptive and, therefore, cannot account for \p mesons produced during the FSI.

\subsubsection{Invariant mass spectrum}

In figure \ref{msigma_phi} we show the mass differential cross section for inclusive $K^+K^-$--photoproduction from $^{40}$Ca. In this calculation the \p is always produced corresponding to its vacuum spectral function, i. e. no medium--modifications of the $\phi$ were considered. In this figure also the density at which the $K^+K^-$--pair was created, i. e. where the \p decayed, is indicated (see figure caption for details). 

In the upper spectrum we included no momentum cut. The relative amount of events where the \p decays at finite nuclear matter density is vanishingly small. Introducing a momentum cutoff (lower spectra) we find an increasing contribution of in--medium decays. Despite the rather strong absorption of the final state particles one gets a fair amount of decays also from higher nuclear densities $\rho>0.5\rho_0$ when a cutoff for the \p momentum of 150 MeV is applied. This should lead to observable consequences if the \p properties in the nuclear environment are indeed modified.

In these spectra one can see very nicely that the mass spectrum will not be distorted by elastic $K^+$-- and $K^-$--FSI. Even the events stemming from \p decays at finite densities form a spectrum which is hardly broader than the free $\phi$--width. This is clearly what one expects due to the following reasons: First, these events still stem from the nuclear surface and therefore the kaons travel only a short distance through a region with finite density. Hence the probability to scatter elastically is very small. Second, these kaons have a very low momentum and therefore the momentum transfer from such a kaon to a nucleon would be very small. Hence, any such scattering processes are hindered by Pauli--blocking.

One further important result of this calculation is the signal to background ratio. From the momentum spectrum in figure \ref{psigma_fsi} it is not clear if the \p signal can easily be distinguished from the non--resonant background if momentum cuts are applied. For small $K^+K^-$--momenta the momentum differential background cross section is much larger than the resonant contribution. Due to the fact that the \p is a very narrow resonance and we therfore look only on a very limited region in $M$ the \p survives as a clearly visible peak, as can be seen in figure \ref{msigma_phi}, bottom.

\subsubsection{Collision broadening}

In figure \ref{msigma_width} we show our results on the $K^+K^-$ invariant mass distribution now including the collision width of the \p following equation (\ref{collwidth}). To compare the width of the obtained spectra with the free spectral function of the \p we fit a Lorentz--curve to the calculated cross sections and compare the fitted curve to a Lorentzian with the free \p width of 4.4 MeV. One can see very nicely how the width increases as the momentum cutoff parameter is decreased. At a cutoff of $\Lambda=100$ MeV we find a total width of the spectrum of 8.2 MeV, which is nearly twice the free width of the $\phi$, corresponding to an average decay density of only $\rho_0/7$. In principle this should be a measurable modification of the spectrum compared to the spectrum obtained from \p photoproduction off the proton.

These results may now be compared to the results reported in \cite{oset}. The authors of \cite{oset} took only the exclusive \p photoproduction mechanism into account. In addition, in their model the FSI are treated as purely absorptive by a $K^-$--absorption factor. On the other hand, these authors included an in--medium change of the $\phi$ properties going beyond the collision width, accounting also for a modifcation of the $K^+/K^-$ spectral functions in the nuclear medium. Despite these obvious differences of the two models the results for the width are in quite good agreement. The reason for this is rather obvious: Due to the large absorption cross sections for the final state particles the calculated spectra are merely sensitive to the nuclear surface. Therefore the production mechanism -- if of primary or secondary nature -- has basically no influence on the observables apart from the overall strength of the cross sections.

\subsubsection{Dropping \p meson mass}

According to Brown--Rho scaling \cite{brown}, the QCD sum rule predictions of Hatsuda and Lee \cite{hatsuda} and the model of \cite{kampfer} we model the \p meson mass shift by introducing a scalar potential $U_S(\vec r)$:
\bea
U_S(\vec r)=-\alpha m_0\frac{\rho(\vec r)}{\rho_0},
\eea
where $\rho(\vec r)$ is the nuclear density, $\rho_0=0.168~\mathrm{fm}^{-3}$, $m_0$ the \p meson pole mass and $\alpha=0.03$. The effective pole mass of a \p in the nuclear environment then is given by
\bea
m_0^*=m_0+U_S(\vec r).
\eea
This potential then corresponds to a downward shift of the \p pole mass of 30 MeV at saturation density.

In figure \ref{msigma_mass} we compare our results on the $K^+K^-$ mass distribution including the mass shift of the \p with the results obtained without the \p mass shift, in both cases including the \p collision width. Since even applying severe momentum cuts most of the $\phi$'s decay in vacuum, the largest contribution to the invariant mass spectra comes from $\phi$ meson decays around the \p vacuum pole mass of 1.02 GeV. The contributions from $\phi$ decays at finite densities and therefore around lower pole masses, lead only to a small shift of spectral strength down to lower invariant masses. Hence, concerning the shape and the position of the pole, we find only tiny modifications of the $K^+K^-$ mass spectrum due to the dropping \p meson mass. Most likely, this effect can experimentally not be distinguished from the mere collision broadening.

On the other hand, the total cross section increases considerably. This increase cannot be explained by the larger phase--space for the production of \p mesons with lower mass, which gives only a five percent correction to the cross section. Instead this rise is due to the weaker absorption of the $K^-$. At saturation density \p mesons are produced around a pole mass of 990 MeV, which lies slightly below the two kaon threshold (992 MeV). These \p mesons are stable with respect to the $K^+K^-$--decay until they travel to the nuclear surface, where their mass again rises due to the smaller nuclear density there. In the nuclear surface the probability of absorption is much smaller and therefore more $K^+K^-$--pairs survive. However, this increase is primarily a theoretical statement and cannot be judged experimentally due to the uncertainties of side--feeding and the cross sections implemented.


\subsection{Kaon potentials}

\subsubsection{Electromagnetic potential}

\label{coulomb}

So far our calculations have been done without taking the Coulomb potential into account. Because of the much larger range of the Coulomb potential compared to the hadronic potentials, transport calculations including the electromagnetic force are connected with a considerable higher computational effort. Nevertheless the effects of the Coulomb potential on the production of low--momentum $K^+K^-$ pairs in nuclei are in principle not negligible.

Since the single--particle energies are constants of motion, the absolut values of the $K^+/K^-$ three momenta in the vacuum far away from the nucleus are given by
\bea
\label{pstar}
|\vec p_{\pm}|=\sqrt{\left(\sqrt{{p^*_{\pm}}^2+m_K^2}\pm V(\vec r_{\mathrm{cr}})\right)^2-m_K^2},
\eea 
whereas $p^*_{\pm}$ are the $K^+/K^-$ three momenta at their creation point $\vec r_{\mathrm{cr}}$, $V(\vec r_{\mathrm{cr}})$ is the absoult value of the Coulomb potential and  $m_K=0.496$ GeV is the kaon mass. The modification of the invariant mass of the $K^+K^-$ pair can be written as
\bea
\Delta M_{\mathrm{inv}}=\sqrt{s}-\sqrt{s^*}\approx\frac{s-s^*}{2M_{\phi}},
\eea
where $\sqrt{s^*}$ denotes the invariant mass of the pair at its creation point and $\sqrt{s}$ the asymptotic value of the $K^+K^-$ invariant mass in the vacuum. Considering a $\phi$ meson decaying at rest in the center of a $^{40}$Ca nucleus, whose Coulomb potential has a maximum depth of $V_0\approx11$ MeV, we find
\bea
\Delta M_{\mathrm{inv}}\approx -\frac{(\vec p_++\vec p_-)^2}{2M_{\phi}}\approx\frac{p_0^2}{M_{\phi}}\left(\sqrt{1-4\frac{(p_0^2+m^2)V_0^2}{p_0^4}}-1\right),
\eea
where $p_0=119$ MeV is the kaon three momentum after the \p decay, yielding a downward shift of the invariant mass of the $K^+K^-$ pair of about 5 MeV. Since this effect is of the same magnitude as the broadening caused by the \p collision width, the Coulomb potential has in principle to be taken into account in our considerations.

Therefore we also have performed full calculations which take account of the Coulomb potential, which is implemented in our model as reported in \cite{Teis1997}. In order to limit the computational effort we run our transport code merely for about two times the life--time of the $\phi$. Finally we determine the asymptotic momenta of the kaons using analytical solutions for Coulomb trajectories.

In figure \ref{coul_corr} we show the results of such a calculation in which -- for further simplification -- we included only the exclusive elementary \p production process, involving no medium modifications of the $\phi$. We find an apparent decrease of the cross section when the Coulomb potential is included. This is due to the fact that the total energy of a $K^-$ can be smaller than the free kaon mass and therefore the expression under the square root of equation (\ref{pstar}) becomes negative. Hence, such a $K^-$ cannot escape from the nucleus, finally leading to a reduction of the total number of $K^+K^-$--pairs in the data sample.

In figure \ref{coul_width} we compare the $K^+K^-$ mass spectra obtained with and without the \p collision width following equation (\ref{collwidth}), in both cases including the Coulomb potential. Despite the clearly visible effects of the collision width when the Coulomb potential was neglected, we find almost no effect due to the \p in--medium width when the potential is included. This can be explained by the fact, that exactly these events which yield the broadening of the spectrum, i. e. kaon pairs stemming from \p decays at finite densities, are reduced by the capture of the negative charged anti--kaons as explained above. Thus, the moderate sensitivity of the mass spectrum to the in--medium width of the $\phi$, which we found in section \ref{phiproduction}, completely vanishes due to the influence of the Coulomb potential.

We, therefore, come to the conclusion, that it is not possible to observe effects of \p medium modifications through the $K^+K^-$ invariant mass distribution.

\subsubsection{Hadronic potentials}

In the nuclear environment the kaons are supposed to feel strongly attractive scalar potentials due to the $KN$--sigma term \cite{kaplan}. In addition there exists also a vector--type interaction which acts repulsively for the kaons and further attractively for the anti--kaons. Hence, one is led to the conlusion that anti--kaons should feel a strong attraction whereas the kaons might feel a slight repulsion, resulting in a moderate increase of the $K$ mass and a substantial downward shift of the $\bar K$ mass with increasing baryon density \cite{schaffner,waas,lutz,ramosk,tolos,schaffnerbielich}.

We explore the implications of the $K^+/K^-$ mass shifts on \p photoproduction from nuclei by introducing the following scalar potentials just in the same way as previously for the \p meson:  
\bea
U_{K^+}(\vec r) & = & +0.08\cdot m_{K^+}^0\frac{\rho(\vec r)}{\rho_0}\label{kpot}\\
U_{K^-}(\vec r) & = & -0.22\cdot m_{K^-}^0\frac{\rho(\vec r)}{\rho_0},\label{kbarpot}
\eea
where $m_{K^+}^0$ and $m_{K^-}^0$ are the vacuum masses of $K^+$ and $K^-$, respectivily, $\rho(\vec r)$ is the nuclear density and $\rho_0=0.168~\mathrm{fm}^{-3}$. For the strength of the potentials we use values obtained by a relativistic mean--field approach in \cite{schaffner}. The effective masses then are given by
\bea
m_K^*=m_K^0+U_K(\vec r).
\eea

Similar to the effects of the Coulomb potential also the scalar kaon potentials modify the $K^+K^-$ invariant mass distribution in $\gamma A$--reactions. Considering, for instance, the momentum of a $K^-$ which is produced inside the nucleus, this momentum has to decrease as this $K^-$ propagates into the vacuum, since the $K^-$ mass increases according to equation (\ref{kbarpot}). The absolut values of the $K^+/K^-$ three momenta in vacuum are given by
\bea
\label{pkstar}
|\vec p_{\pm}|=\sqrt{{\vec p_{\pm}^{~*\,2}}+{m_{K^+/K^-}^{*\,2}}-m_K^2},
\eea
where $\vec p_{\pm}^{~*}$ and $m_{K^+/K^-}^*$ are the momenta and effective masses of $K^+$ and $K^-$ in--medium and $m_K$ is the kaon vacuum mass.

In figure \ref{msigma_kpot} we show our results including the above introduced kaon potentials in comparison to a calculation without kaon potentials, in both cases including the \p collision width. The most obvious difference is a considerable decrease of the cross sections when the potentials are included. This decrease is due to the same effect as in the electromagnetic case: The total energy of a $K^-$ which is produced with low momentum inside the nucleus might be smaller than the $K^-$ vacuum mass. Therefore such a $K^-$ cannot escape from the nucleus and finally will be absorbed in channels like $K^-N\rightarrow \pi\Lambda$.

The second important observation is the shift of spectral strength to higher invariant masses. This behaviour one can understand as follows: The invariant mass squared of a $K^+K^-$--pair can be expressed in the following way:
\begin{eqnarray}
s=(E_++E_-)^2-|\vec p_+|^2-|\vec p_-|^2-2|\vec p_+||\vec p_-|\cos \theta,
\end{eqnarray}
where $E_+,\vec p_+$ and $E_-,\vec p_-$ are the energies and momenta of the $K^+$ and the $K^-$, respectively, and $\theta$ is the angle between their three momenta. At this point we consider for simplicity only a mass modification of the $K^-$, whereas the $K^+$ properties remain unchanged. Since the absolut value of the $K^-$ three momentum decreases as it propagates from the interior of a nucleus to the vacuum, we can write $|\vec p_-|=|\vec p_-^{~*}|-\Delta p$ with $\Delta p\ge 0$, given by equations (\ref{kbarpot}) to (\ref{pkstar}). Assuming for simplicity that the directions of the kaon three momenta remain unchanged, the modification of the invariant mass squared of the pair then can be written as
\begin{eqnarray}
\label{invmass}
s-s^*=\Delta p\left(|\vec p_-|+|\vec p_-^{~*}|+2(p_{\phi}^{2}-p_{\mathrm{cm}}^{2})/|\vec p_-^{~*}| \right),
\end{eqnarray}
where $s^*$ is the invariant energy squared of the pair at its creation point inside the nucleus and $s$ is the invariant energy squared after the propagation to the vacuum. $p_{\phi}$ denotes the absolute value of the \p three momentum in the nuclear rest frame and $p_{\mathrm{cm}}$ is the absolute value of the kaon three momentum in the \p rest frame. 

If the $\phi$ is at rest inside the nucleus, i. e. $p_{\phi}=0$, then $|\vec p_+^{~*}|=|\vec p_-^{~*}|=p_{\mathrm{cm}}$. We therefore find $s-s^*=-\left(\Delta p\right)^2\le 0$ and hence the invariant mass after the propagation is lower than the mass of the decaying $\phi$ meson. Considering an increasing momentum of the decaying $\phi$, the expression inside the brackets of equation (\ref{invmass}) at some point becomes positive, anyway 
when $|\vec p_{\phi}|\ge p_{\mathrm{cm}}$. This means that the invariant mass reconstructed from the $K^+K^-$--pair is larger than the mass of the decayed $\phi$ meson. This is exactly what one can observe in figure \ref{msigma_kpot}. Note also that the momentum cutoff $\Lambda$ is applied to the three momenta of the $K^+K^-$--pairs and therefore also $\phi$'s with momenta larger than $\Lambda$ contribute to the spectra due to the decreasing $K^-$ momentum.

Finally, we show in figure \ref{final} the results of a full calculation, taking into account elastic and inelastic photoproduction of \p mesons as well as non--resonant background production of $K^+K^-$--pairs from the nucleon. We further included the kaon potentials given by equations (\ref{kpot}) and (\ref{kbarpot}), the \p in--medium width following equation (\ref{ansatz}), also taking care of the modified phase space for the $K\bar K$--decay, and the Coulomb potential. In this respect figure \ref{final} is our prediction for the experimental cross sections, which will be measured at Spring8 in Osaka \cite{osaka}. For statistical reasons the spectra are folded with a Gaussian with a width of 2 MeV. The measurement of such a broad spectrum with an asymmetry to higher invariant masses as depicted in figure \ref{final} would give further experimental evidence for the modifications of the masses of kaons and anti--kaons at finite baryon density. On the other hand, these spectra can give no information about the \p properties in the nuclear medium.

The effect of the kaon potentials on the propagation of kaons and anti--kaons was not taken into account in \cite{oset}. On the other hand, the \p width used in the model of \cite{oset} respects also a renormalization of the kaon properties in the nuclear medium. Since the authors of \cite{oset} neglected also the influence of the Coulomb potential on the invariant mass spectrum, which they estimated to yield a shift of the invariant mass of less than 1 MeV, they found an observable broadening of the $K^+K^-$ mass distribution due to the $\phi$ in--medium width.


\section{Summary}\label{summary}

We have studied the consequences of medium--modifications of the \p meson as well as of kaons and anti--kaons on the $K^+K^-$ invariant mass distribution in a photonuclear reaction. To this purpose we have exploited the semi--classical BUU transport model, which allows for a realistic treatment of the FSI. We have included in our calculation the elastic elementary reaction, i. e. $\gamma N\rightarrow \phi N$, as well as inelastic \p photoproduction channels, i. e. $\gamma N\rightarrow \phi X, X\ne N$. We have shown that the inelastic channels give the dominant contribution in the interesting region of low \p momenta.

In the FSI we have included all essential $\phi N$ reactions, for which the total cross section was constrained by the VMD relation to high energy \p photoproduction. Due to inelastic scattering processes $\phi N\rightarrow \phi X$ we have found an appreciable enhancement of the yield of low--momentum \p mesons compared to calculations disregarding the FSI. 

Applying severe momentum cuts to the \p three momentum we have found a fair contribution of in--medium decays of the \p to the $K^+K^-$ invariant mass distribution. In a calculation including the collision width of the \p meson we have shown that a decreasing momentum cut leads to an enhancement of collision broadened $\phi$'s in the data sample and hence to an increasing width of the obtained spectra. Applying a cut of 100 MeV we found a total width of the $K^+K^-$ mass distribution of 8.2 MeV which is nearly twice the free \p width. Taking into account also the Coulomb potential we found that the sensitivity of the mass spectrum to the \p in--medium width completely vanishes. Due to the strong absorption of the final state particles we also found no measurable effect caused by a shift of the \p meson pole mass.

In contrast to the vanishing sensitivity of the $K^+K^-$ mass spectrum to the in--medium properties of the \p we have found a rather strong impact of potentials of kaons and anti--kaons on the mass differential cross sections. Due to the propagation of the kaons from their in--medium mass back to their vacuum mass we have obtained an broadened spectrum with an obvious shift of spectral strength to higher invariant masses. This effect definitely should be visible in experiment.

In summary, we state that the \p in--medium properties -- even applying severe momentum cuts -- are not visible through the $K^+K^-$ invariant mass distribution. Nevertheless the considered reaction can give further evidence for the modifiction of the masses of kaons and anti--kaons at finite baryon density.



\bc
\bfi
\bc
\igr[scale=.65]{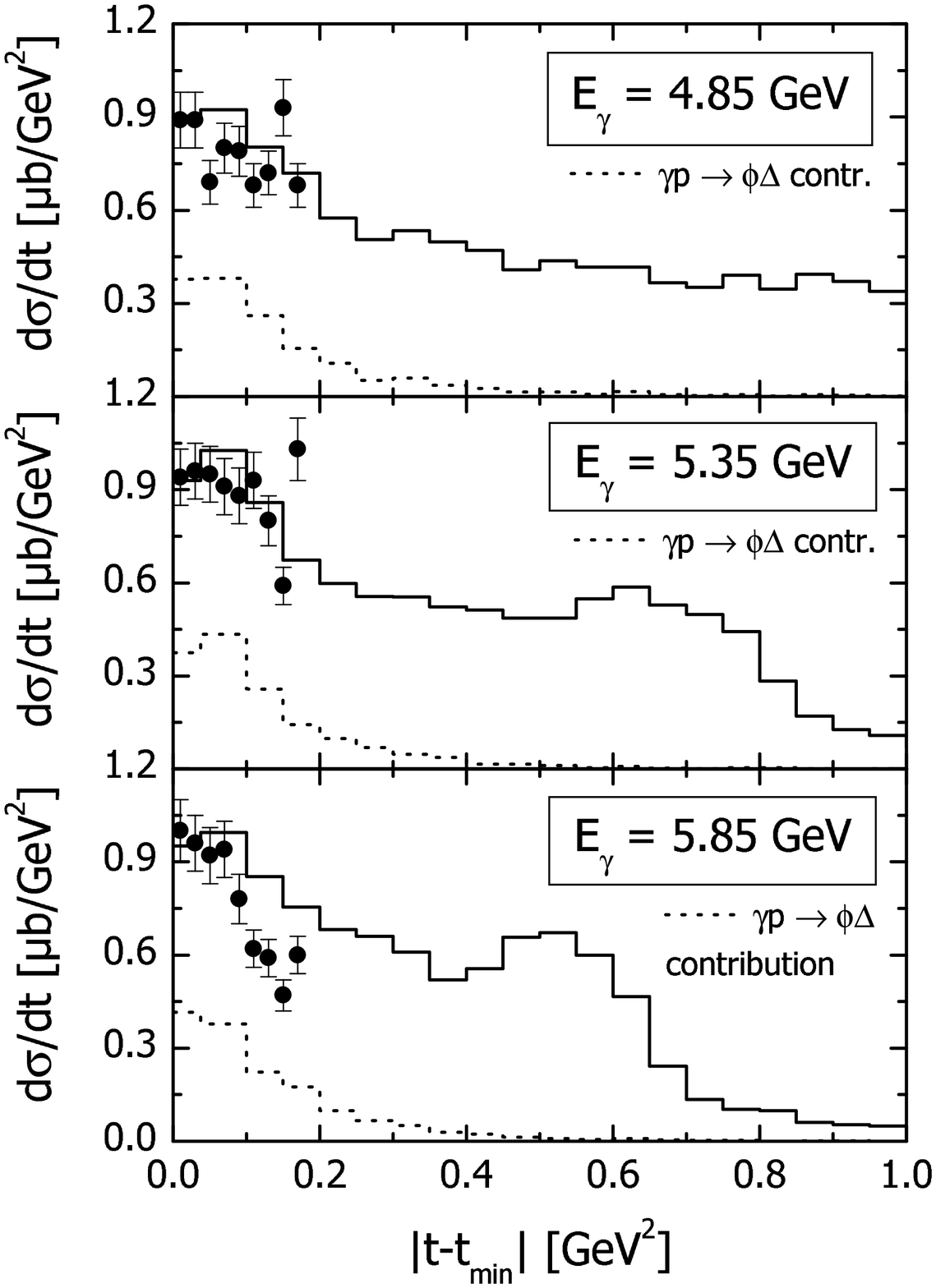}
\caption{Differential cross section for inclusive $\phi$ photoproduction $\gamma p\rightarrow \phi X$, whereas $1.2$ GeV$\le M_X<2.1$ GeV. The experimental data is taken from \cite{desy}.}
\label{tsigma_inc}
\ec
\efi
\ec

\bc
\bfi
\bc
\igr[scale=.55]{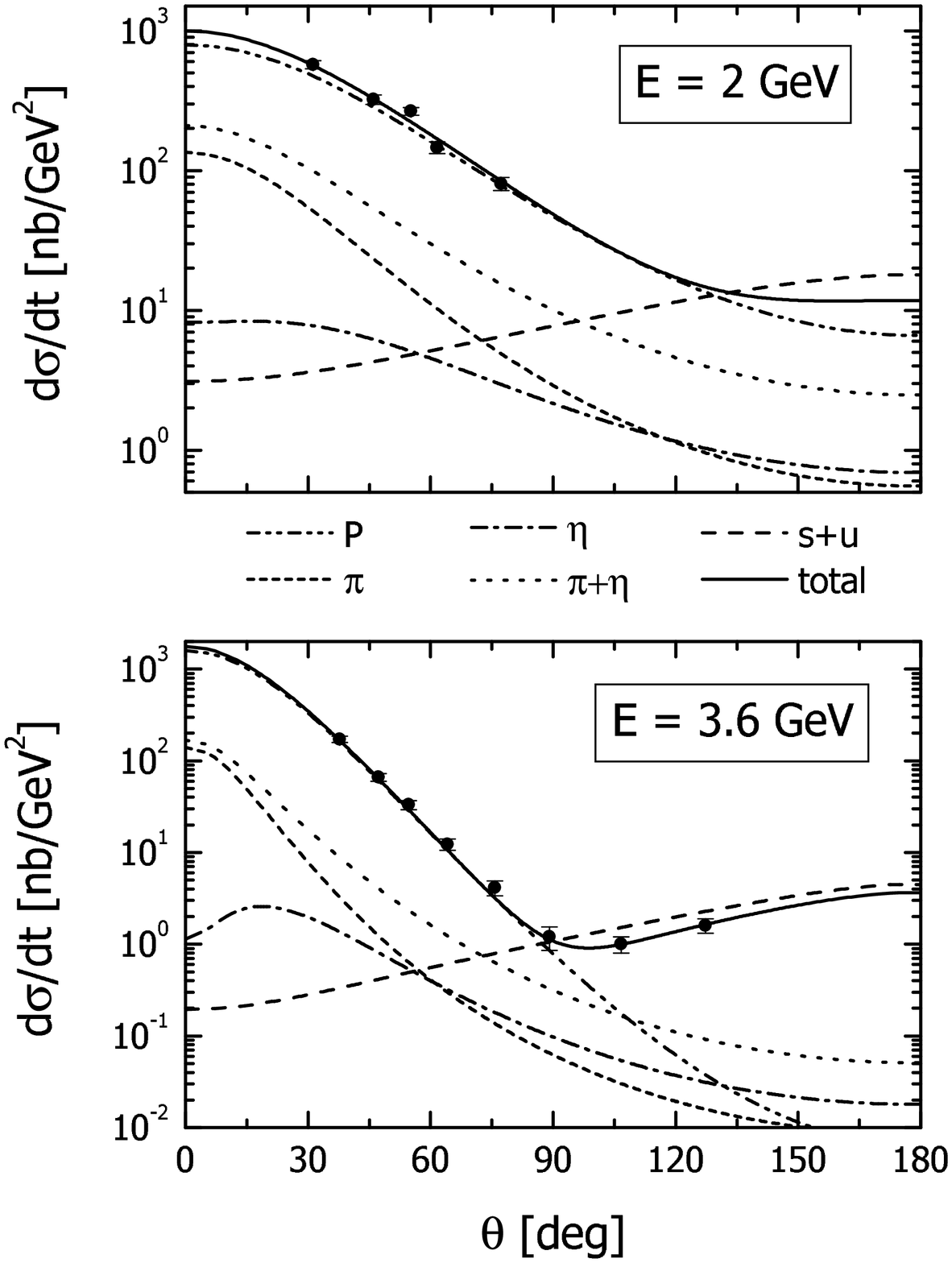}
\caption{Differential cross section for exclusive $\phi$--photoproduction from the proton. For details see section \ref{xsection} and \cite{titov}. The experimental data stem from \cite{be74}(2 GeV) and \cite{ac00}(3.6 GeV).}
\label{tsigma_exc}
\ec
\efi
\ec

\bc
\bfi
\bc
\igr[scale=.5]{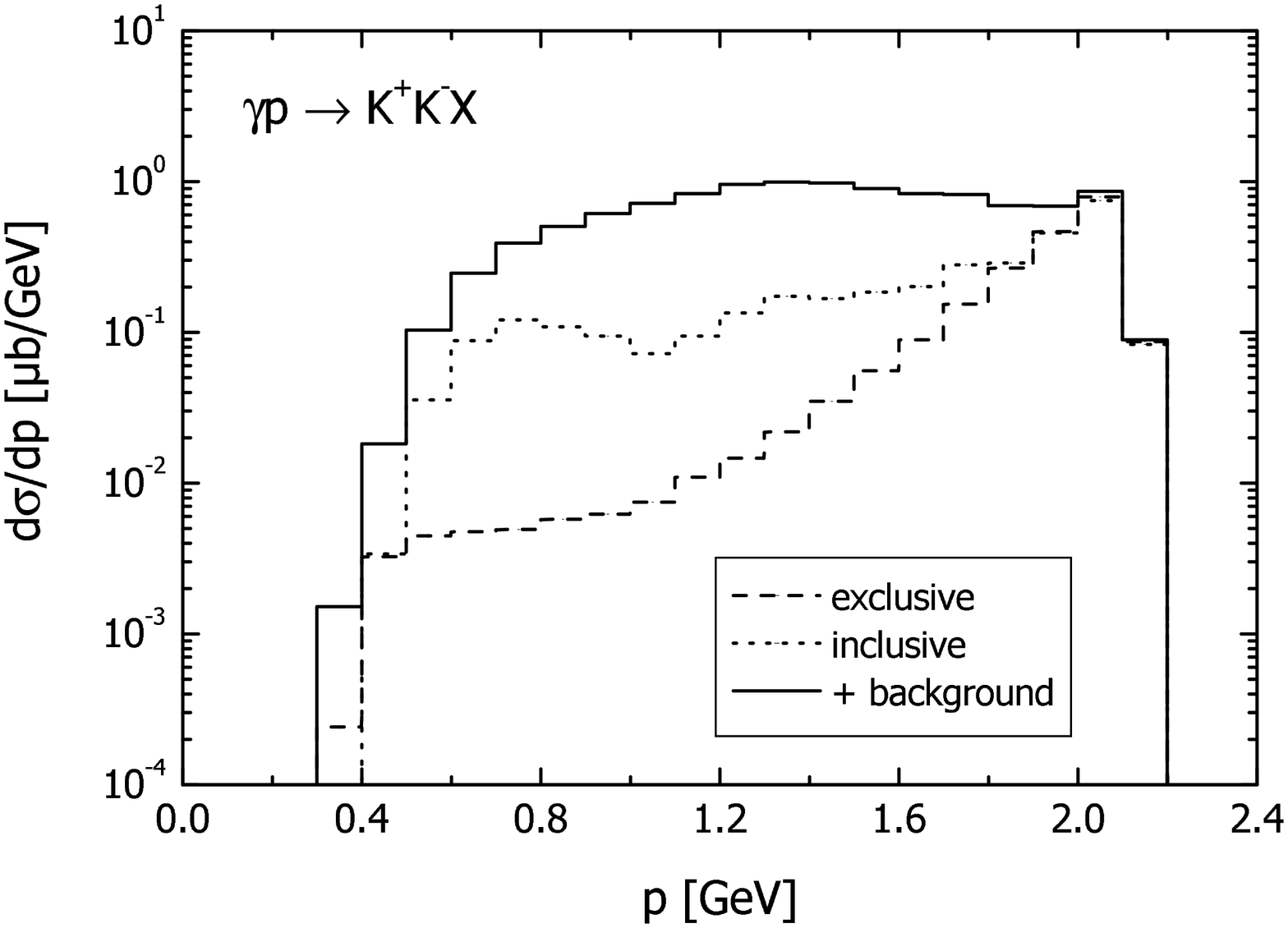}
\caption{Momentum differential cross section for photoproduction of $K^+K^-$--pairs from the proton at 2.4 GeV photon energy. The different curves correspond to the exclusive \p production process $\gamma p\rightarrow \phi p\rightarrow K^+K^-p$, to the inclusive \p production process $\gamma p\rightarrow \phi X\rightarrow K^+K^- X$ and to the total inclusive $K^+K^-$ photoproduction reaction.}
\label{psigma_exinc}
\ec
\efi
\ec

\bc
\bfi
\igr[scale=.5]{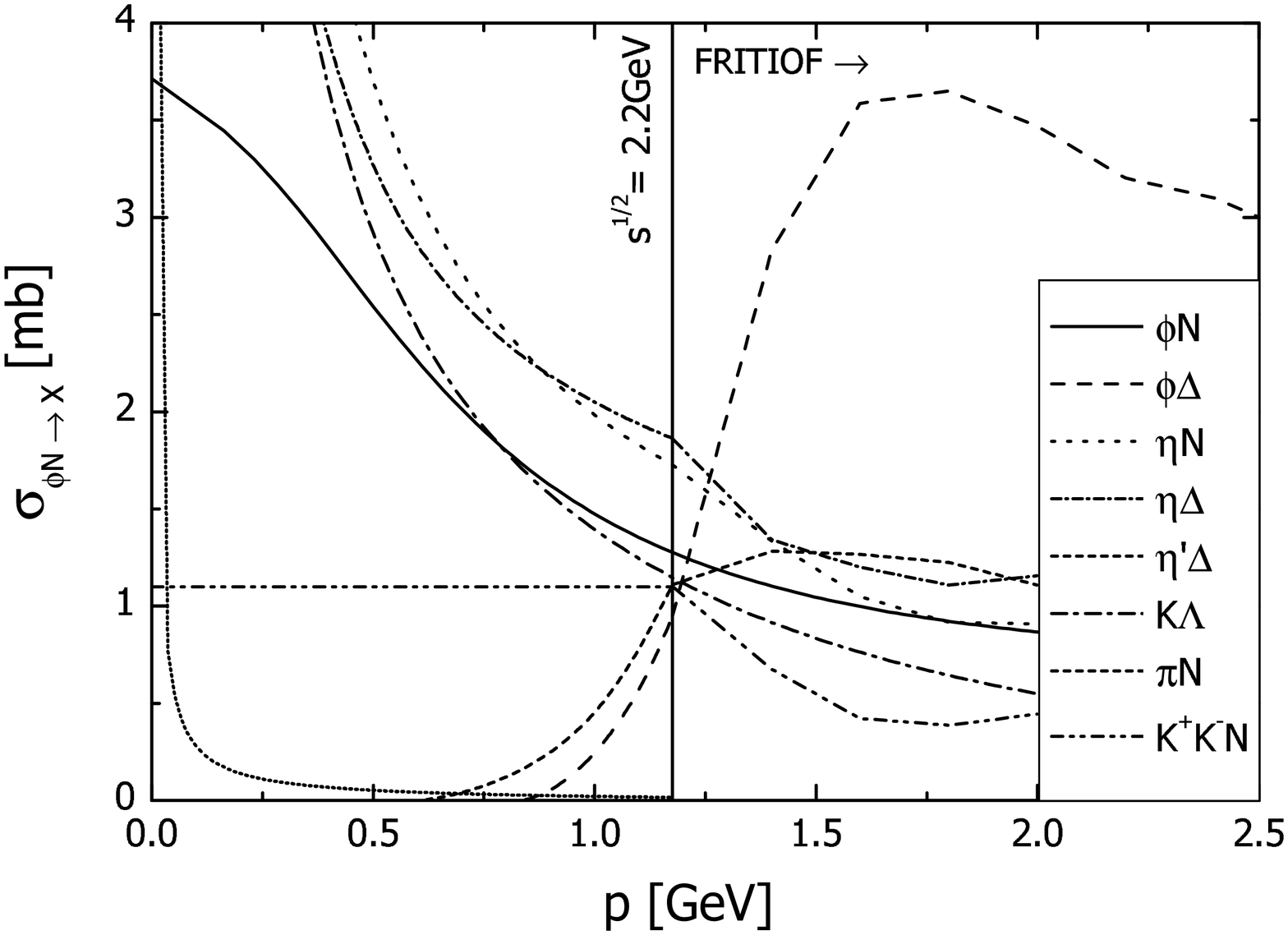}
\caption{Low energy $\phi$--nucleon cross sections. For energies below $\sqrt{s}=2.2$ GeV the plotted cross sections are calculated as described in section \ref{phin}. Apart from the elastic scattering process the cross sections above $\sqrt{s}=2.2$ GeV are calculated with the FRITIOF \cite{fritiof} model, where we put in a total $\phi N$ cross section of 12 mb.}
\label{sigma_low}
\efi
\ec

\bc
\bfi
\bc
\igr[scale=.5]{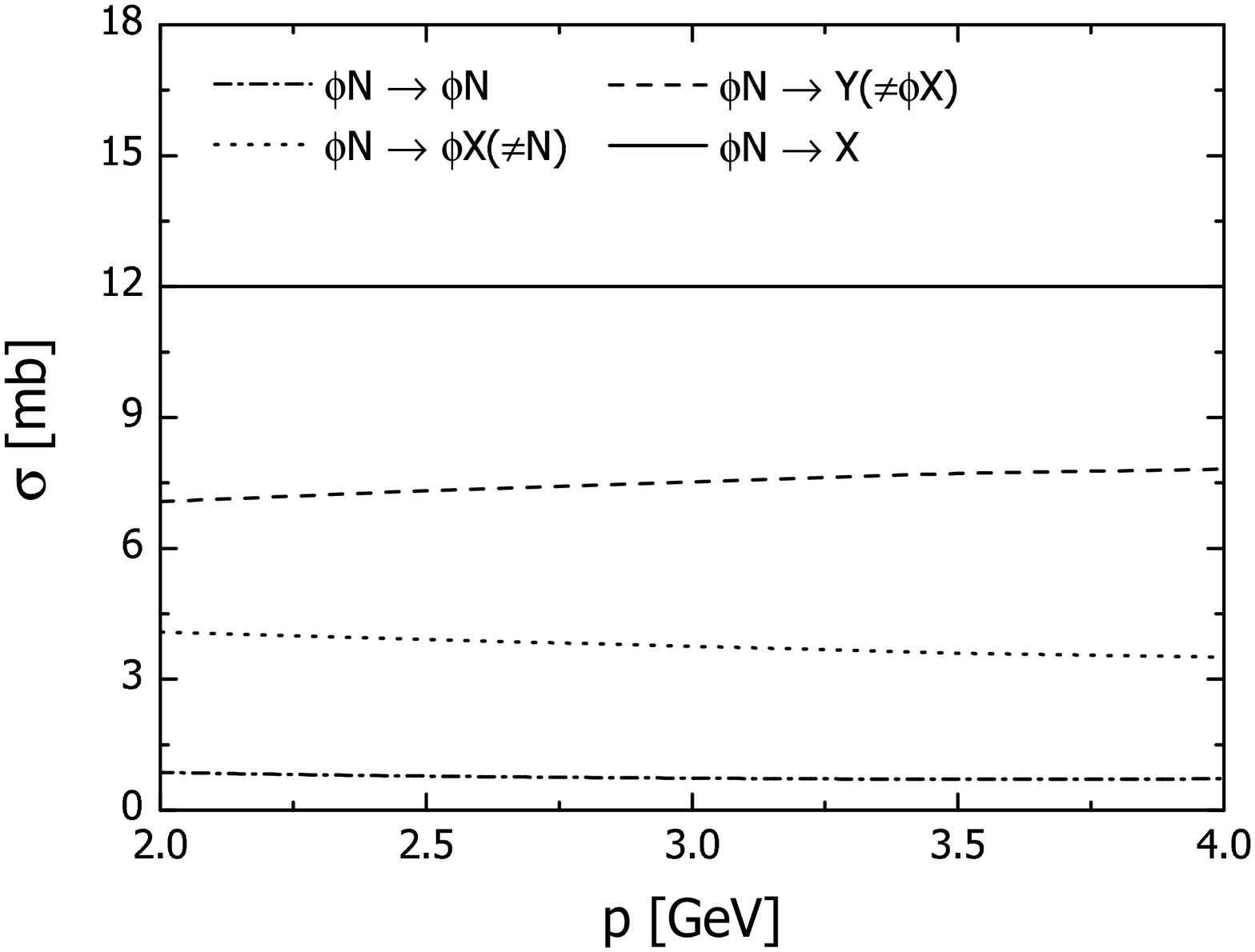}
\caption{High energy cross section for $\phi$--nucleon collisions calculated with FRITIOF. We plotted seperately the cross sections for elastic and inelastic $\phi N$ scattering as well as the cross section for \p absorption on the nucleon.}
\label{sigma_high}
\ec
\efi
\ec

\bc
\bfi
\bc
\igr[scale=.5]{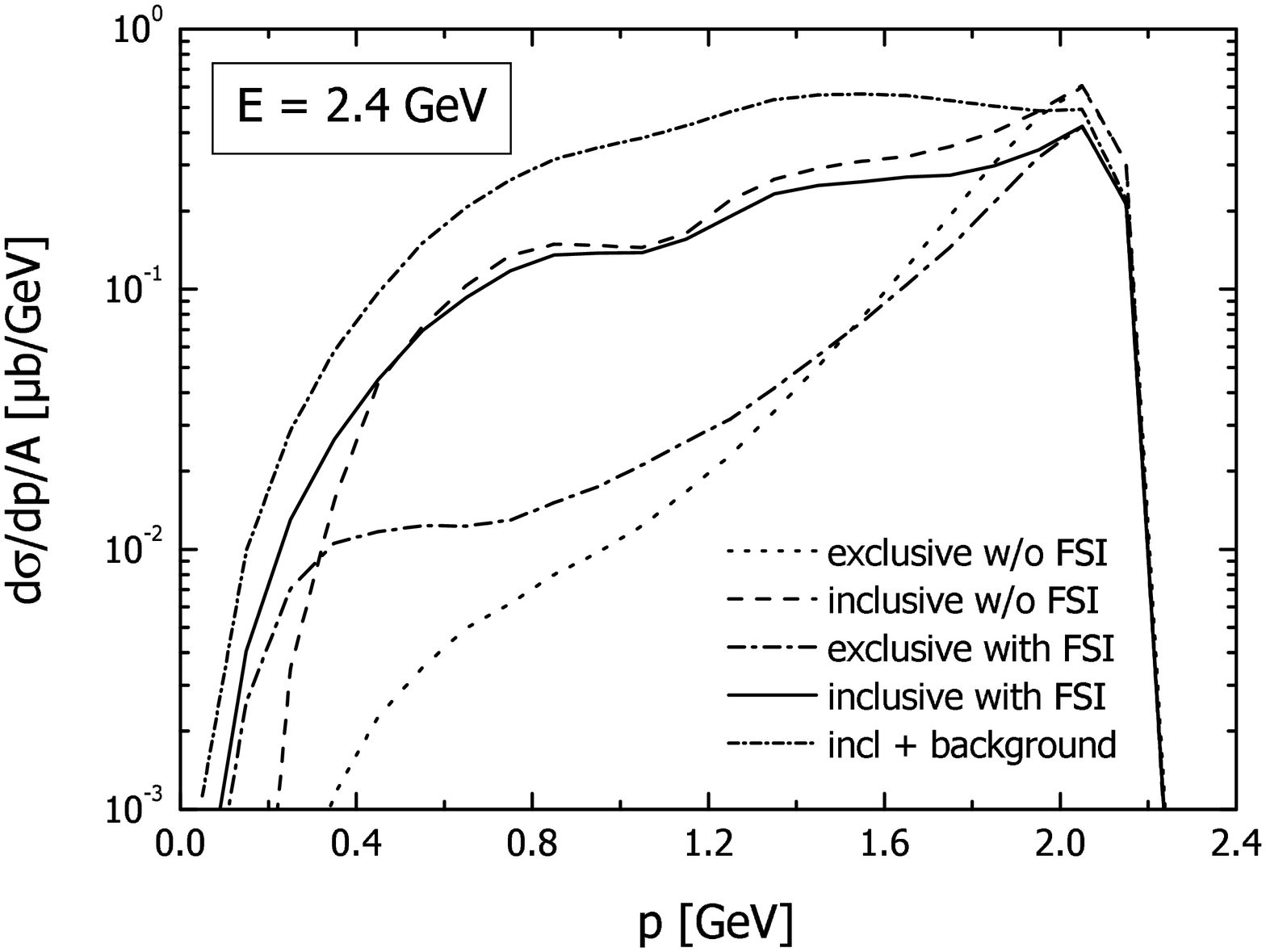}
\caption{Momentum differential cross section for  $K^+K^-$ photoproduction from $^{40}$Ca. We show seperately the contribution from the elastic and inclusive \p production processes. In both cases a calculation including FSI and a calculation without FSI is plotted. Furthermore we show the total inclusive $K^+K^-$ photoproduction cross section including non--resonant background production of $K^+K^-$--pairs.}
\label{psigma_fsi}
\ec
\efi
\ec

\bc
\bfi
\bc
\igr[scale=.5]{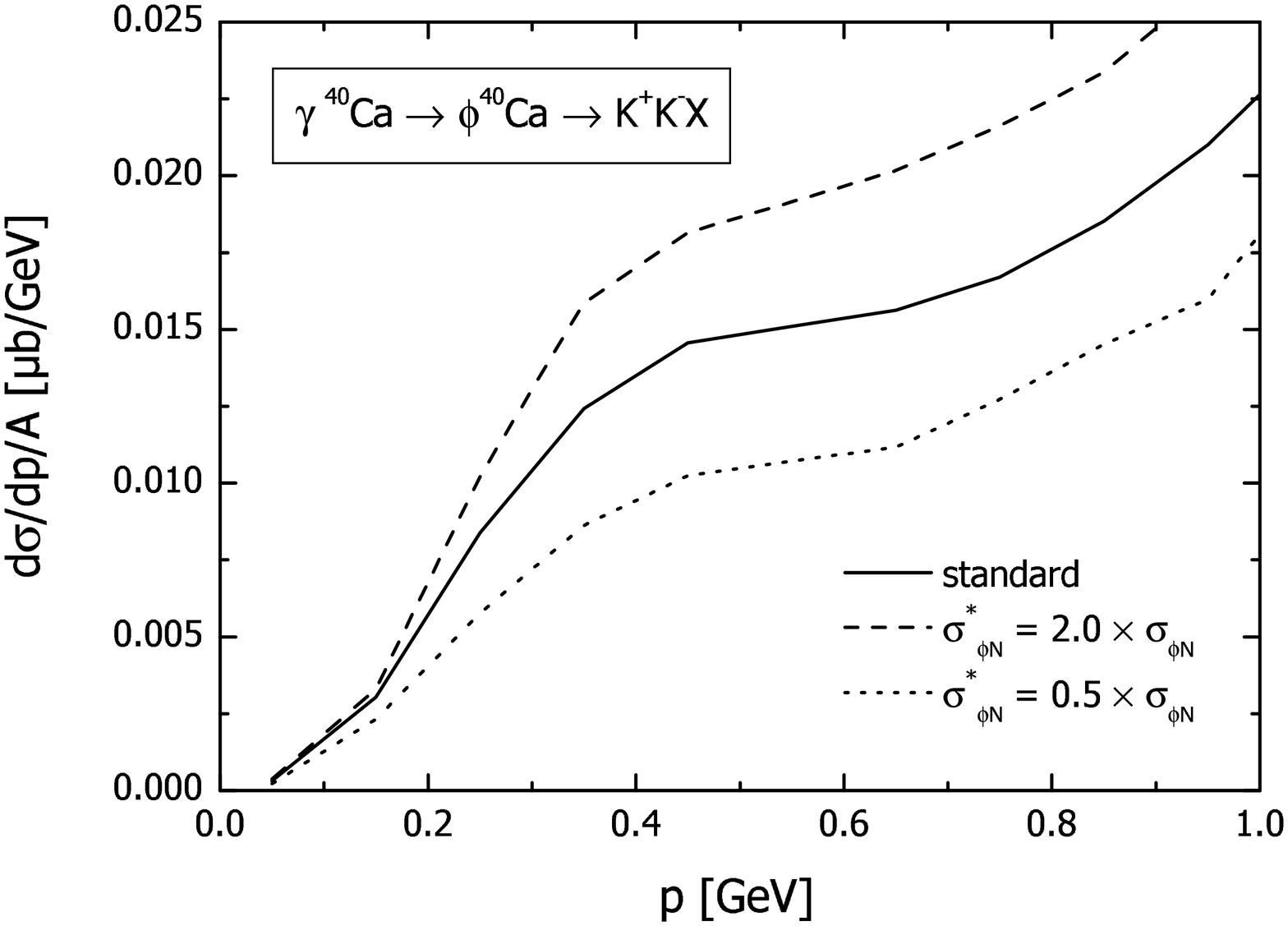}
\caption{Momentum differential cross section for $\phi$ photoproduction from $^{40}$Ca, taking into account only the exclusive elementary reaction at a photon energy of 2.4 GeV. For the dashed(dotted) line the $\phi$--nucleon cross sections are multiplied by a factor of 2(0.5).}
\label{psigma_enh}
\ec
\efi
\ec

\bc
\bfi
\bc
\igr[scale=.65]{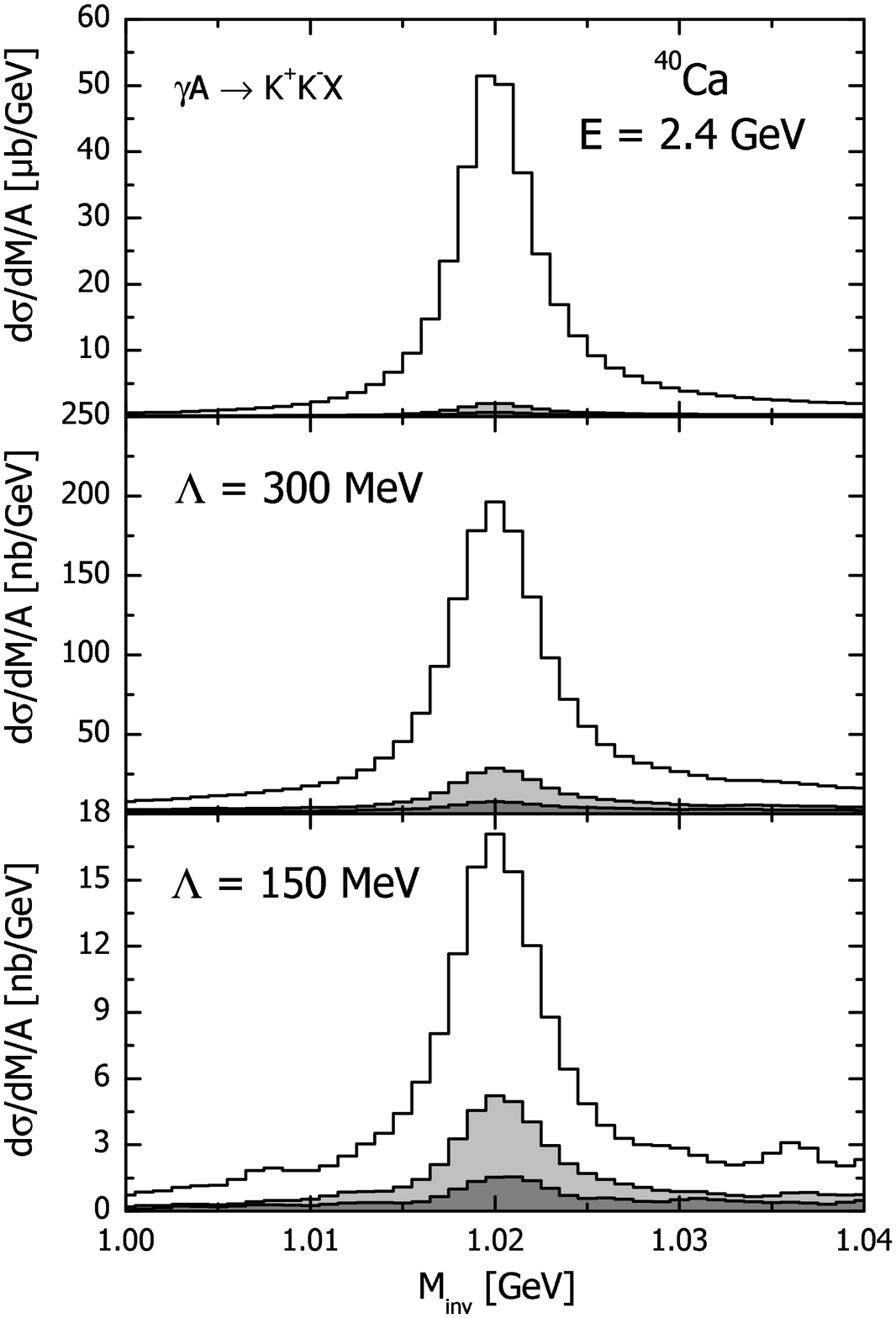}
\caption{Invariant mass distribution of $K^+K^-$--pairs without any medium modifications of the $\phi$. The light grey shaded curves correspond to events where the $K^+K^-$--pair was produced at densities $\rho>0.1\rho_0$, whereas the dark grey shaded curves correspond to creation densities of $\rho>0.5\rho_0$. $\Lambda$ denotes the momentum cutoff parameter.}
\label{msigma_phi}
\ec
\efi
\ec

\bc
\bfi
\bc
\igr[scale=.65]{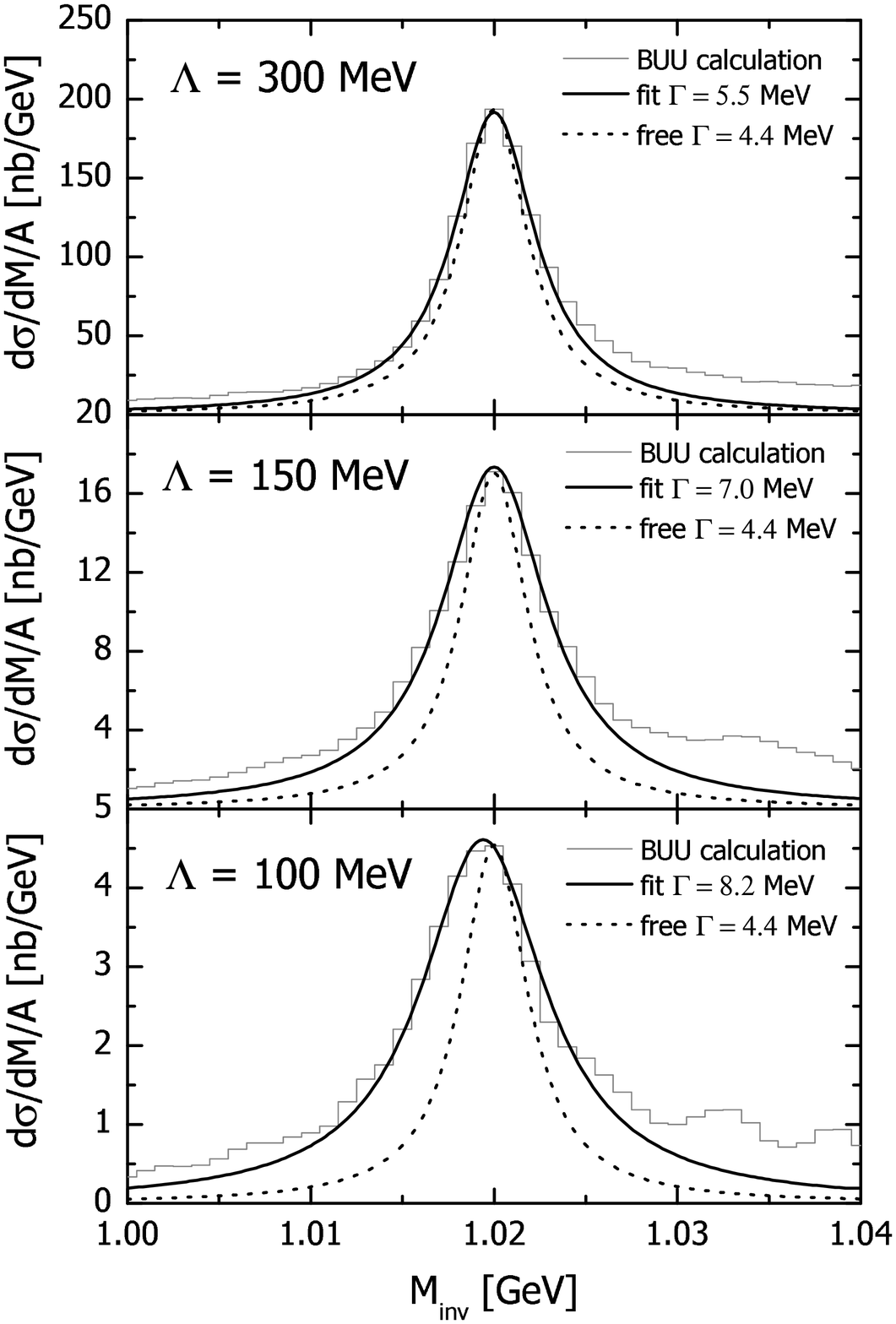}
\caption{$K^+K^-$ mass distribution including a collision width of the $\phi$ meson corresponding to equation \ref{collwidth}. We show a Lorentz--fit to our calculated spectra besides a Lorentzian with the free \p width of 4.4 MeV.}
\label{msigma_width}
\ec
\efi
\ec

\bc
\bfi
\bc
\igr[scale=.65]{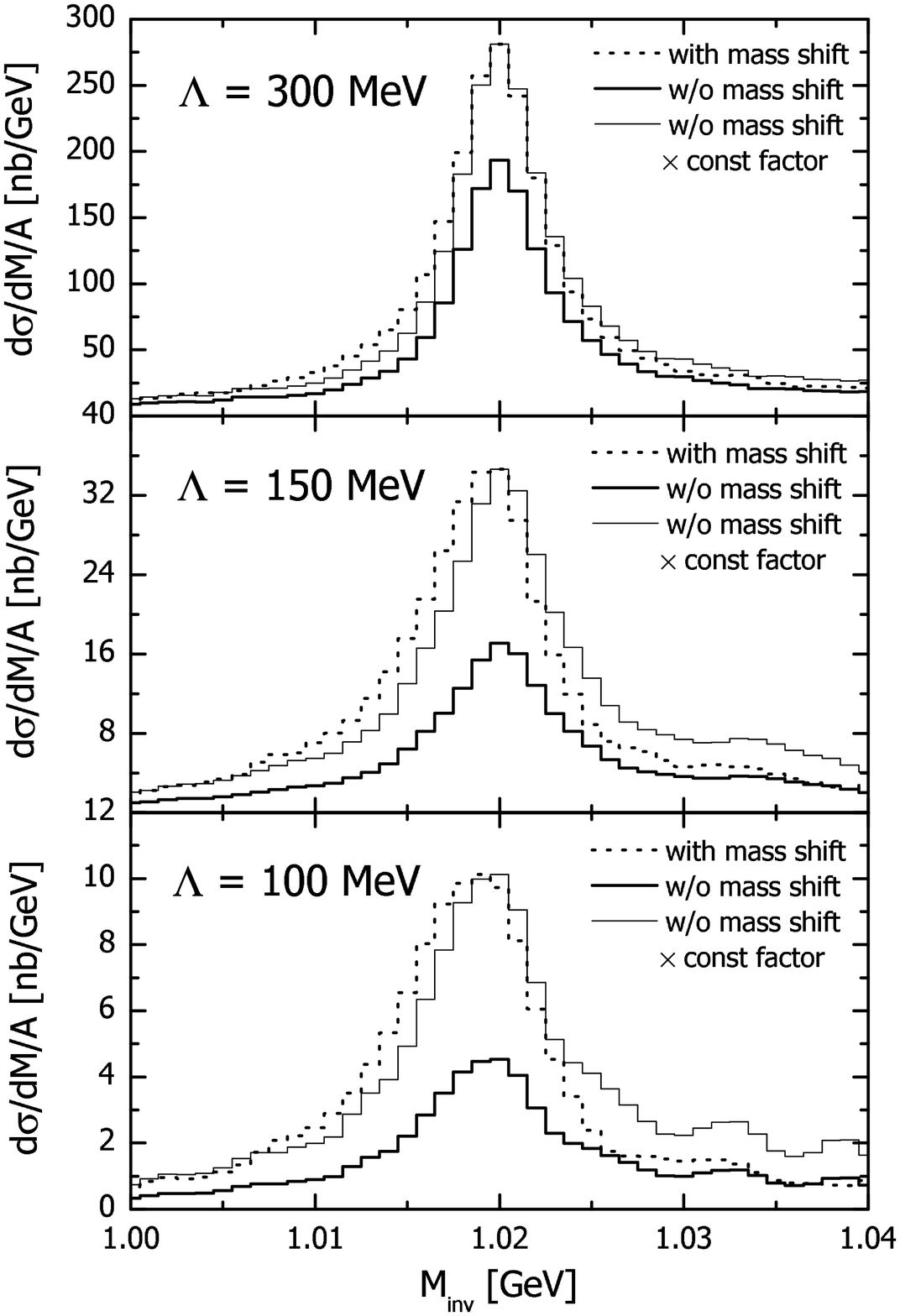}
\caption{Comparison of $K^+K^-$ mass distributions with and without a $\phi$ meson mass shift of 30 MeV at saturation density. The thin solid line corresponds to the same spectrum as the thick solid line merely multiplied by a constant factor in order to compare the shape of the spectrum with the spectrum calculated without the \p mass shift.}
\label{msigma_mass}
\ec
\efi
\ec

\bc
\bfi
\bc
\igr[scale=.5]{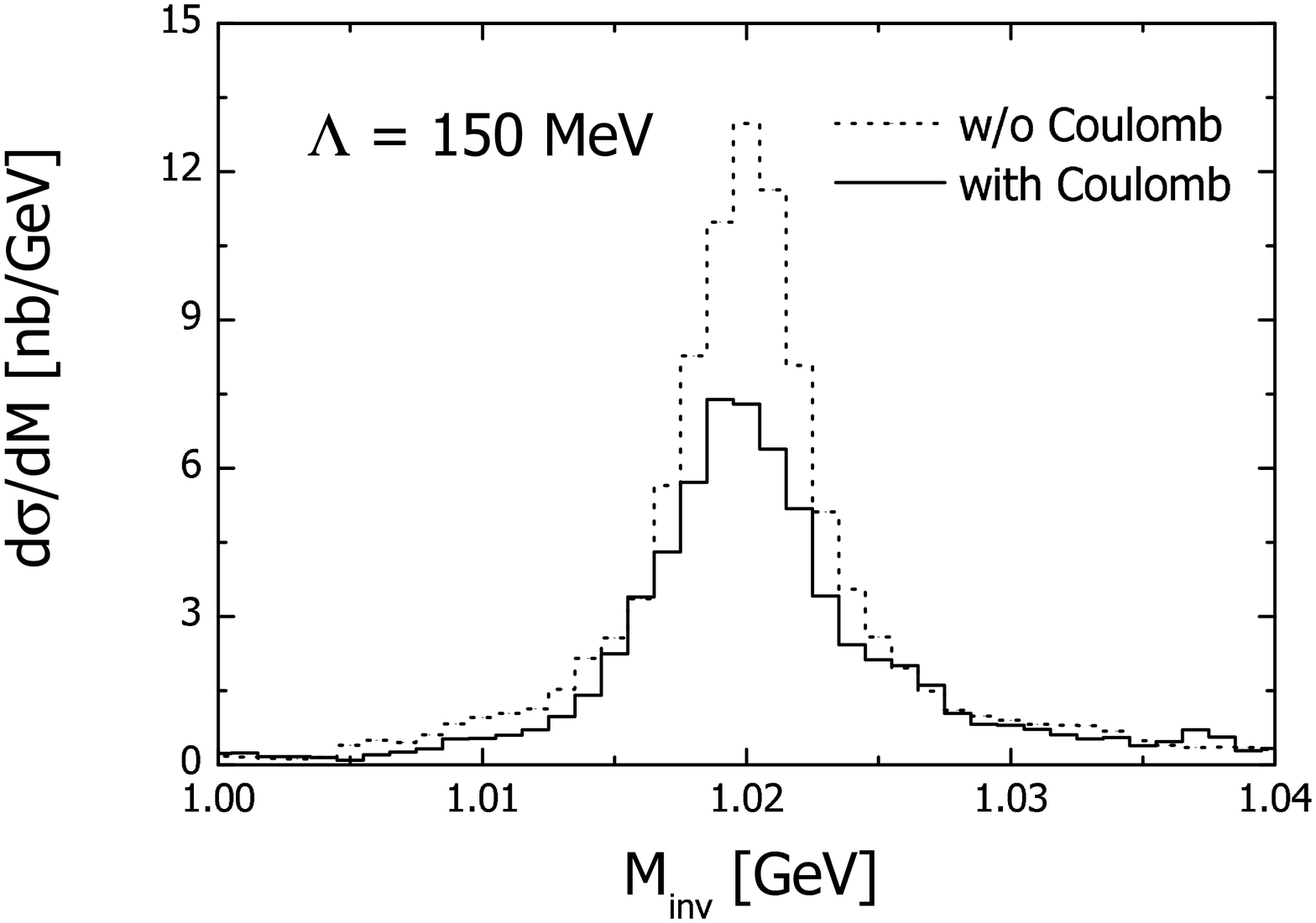}
\caption{$K^+K^-$ mass spectrum of a calculation taking into account only exclusive \p production from the nucleon with a momentum cut of 150 MeV. For the solid line we included the Coulomb potential, see section \ref{coulomb}.}
\label{coul_corr}
\ec
\efi
\ec

\bc
\bfi
\bc
\igr[scale=.5]{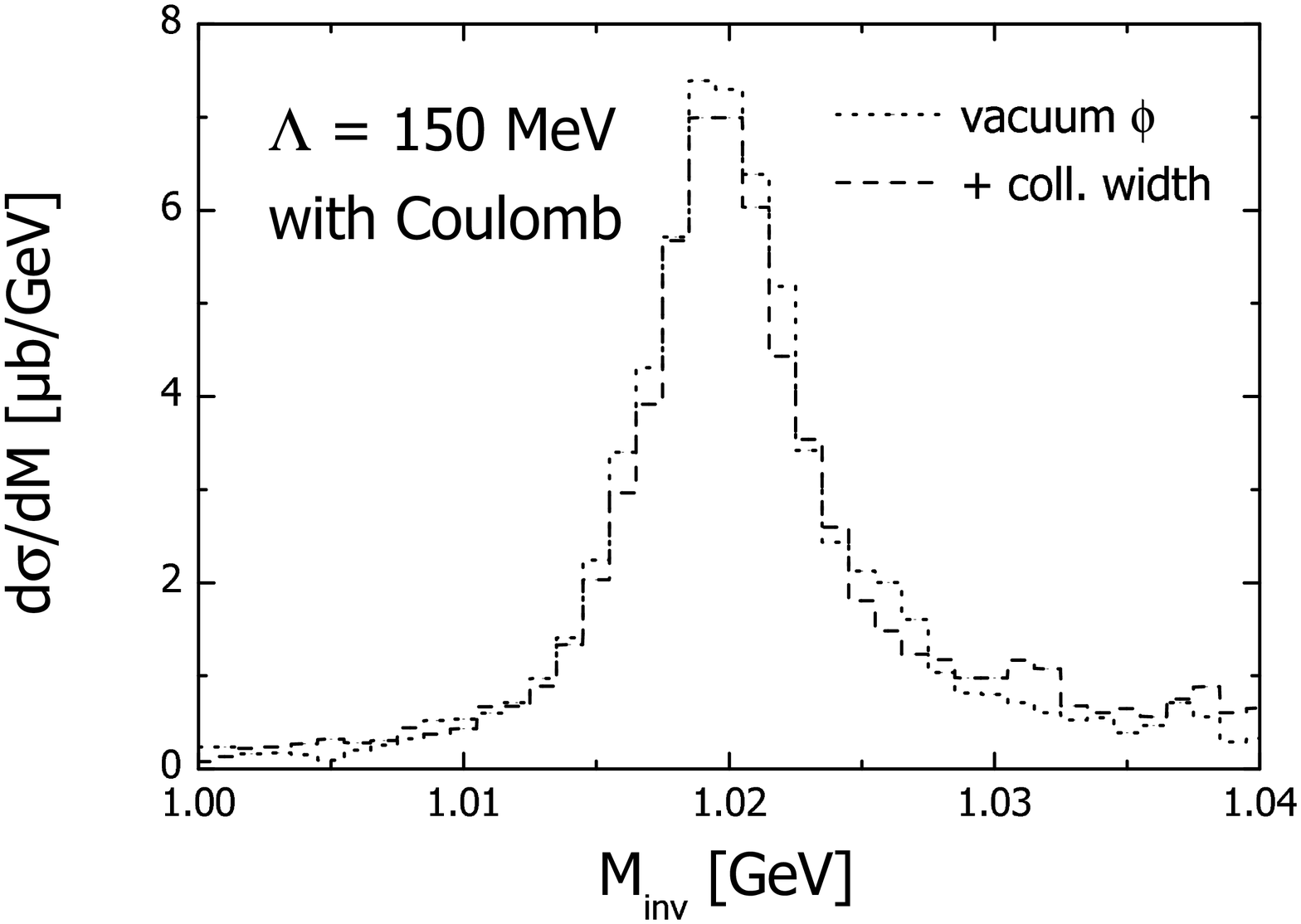}
\caption{Comparison of Coulomb corrected $K^+K^-$ mass spectra of a calculation taking into account only the exclusive \p photoprodcution process with a momentum cut of $\Lambda=150$ MeV. The dotted line is calculated without collision width whereas the dashed line corresponds to a calculation with collision width.}
\label{coul_width}
\ec
\efi
\ec

\bc
\bfi
\bc
\igr[scale=.65]{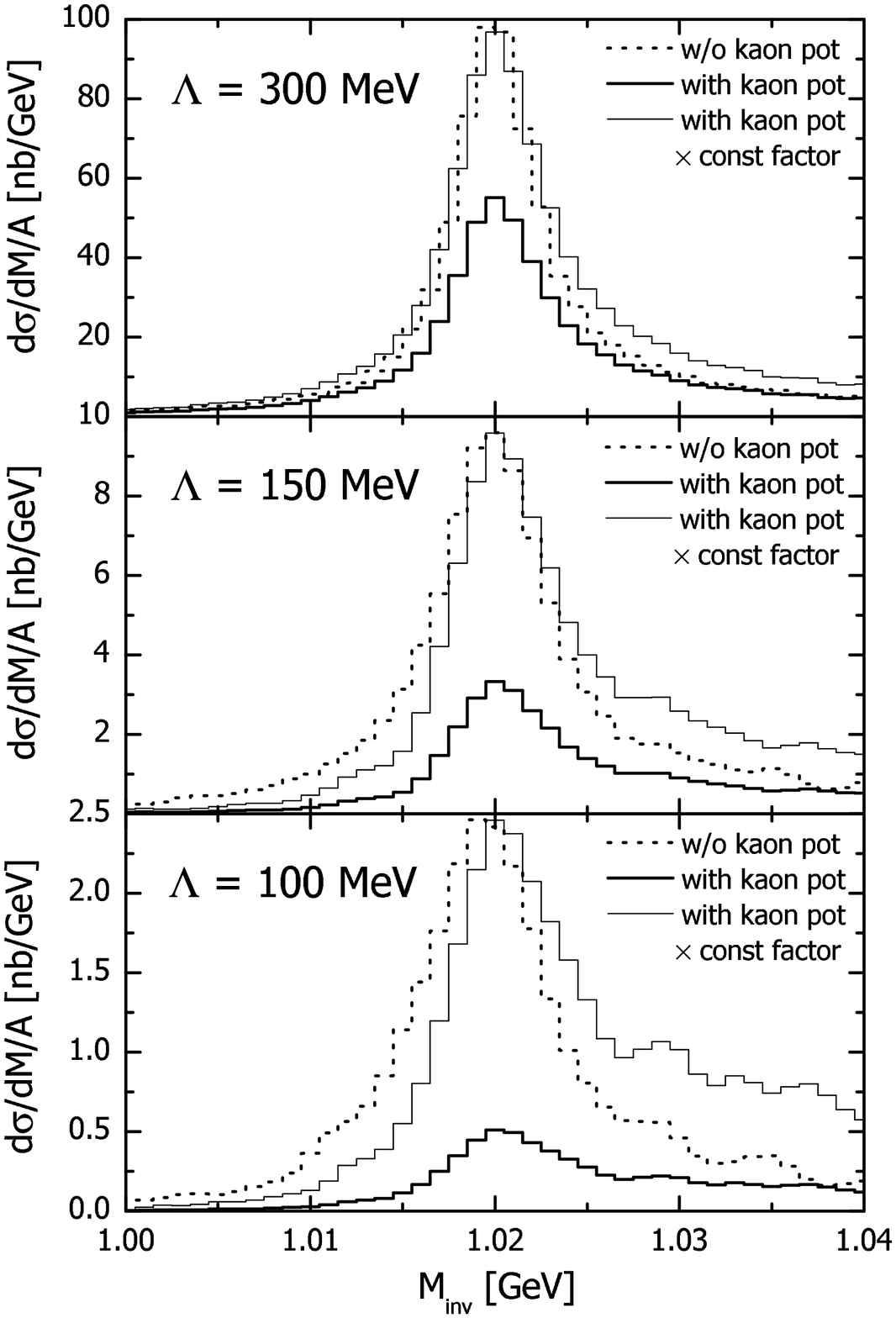}
\caption{Comparison of mass differential cross sections including the collision width of the \p meson calculated with and without $K^+$ and $K^-$ potentials. The thin solid line is equal to the thick solid line times a constant factor.}
\label{msigma_kpot}
\ec
\efi
\ec

\bc
\bfi
\bc
\igr[scale=.5]{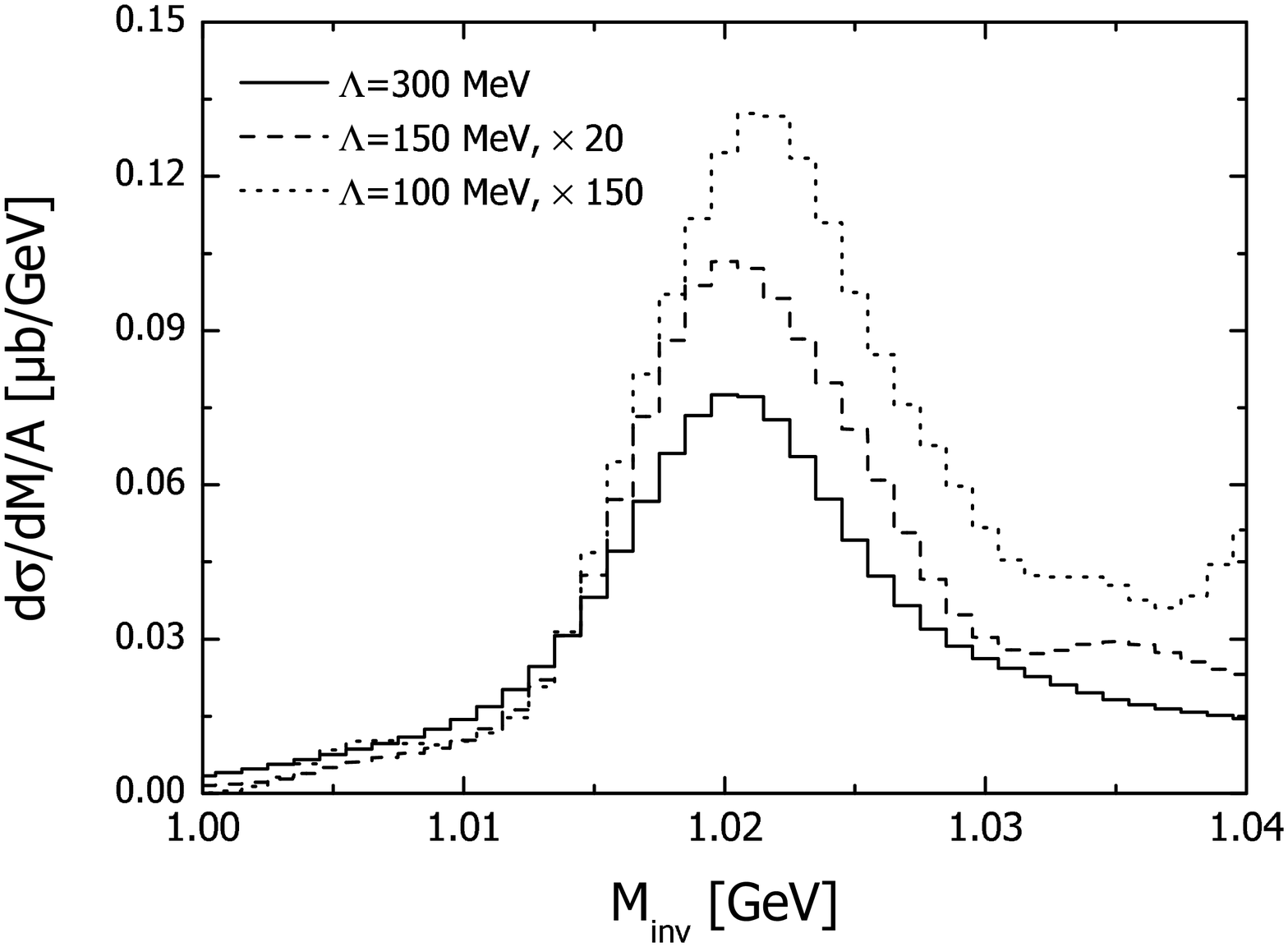}
\caption{Mass differential cross section for inclusive photoproduction of $K^+K^-$--pairs from $^{40}$Ca at a photon energy of 2.4 GeV. The calculation includes elastic and inelastic \p photoproduction as well as non--resonant background production of $K^+K^-$--pairs from the nucleon. We took into account the mass modifications of kaons and anti--kaons, following equations (\ref{kpot}) and (\ref{kbarpot}), as well as the Coulomb potential. The dashed and dotted curves are multiplied by constant factors (see labels).}
\label{final}
\ec
\efi
\ec

\end{document}